
\global\newcount\meqno
\def\eqn#1#2{\xdef#1{(\secsym\the\meqno)}
\global\advance\meqno by1$$#2\eqno#1$$}
%
\global\newcount\refno
\def\ref#1{\xdef#1{[\the\refno]}
\global\advance\refno by1#1}
\global\refno = 1
\vsize=7.5in
\hsize=5in
\magnification=1200
\tolerance 10000
%
%
%
%
\def\dofig{0}
%
%
%
\font\sevenrm = cmr7

\hyphenation{non-over-lapping}

\def\calz{{\cal Z}}
\def\calzbk{{\cal Z}_{\beta,k}}

\def\lbk{\lambda_{\beta,k}}

\def\parmu{\partial_{\mu}}

\def\s#1{{\bf#1}}
\def\sp{ \s p}
\def\sx{\s x}

\def\sumn{\sum_{n=-\infty}^{\infty}}
\def\sumo{\sum_{n=1}^{\infty}}
\def\pmb#1{\setbox0=\hbox{$#1$}%
\kern-.025em\copy0\kern-\wd0
\kern.05em\copy0\kern-\wd0
\kern-.025em\raise.0433em\box0 }
\medskip

\ifnum\dofig=1 
\input epsf
\fi

\baselineskip=0.1cm
\medskip
\nobreak
\medskip

\baselineskip 12pt plus 1pt minus 1pt
\vskip 2in
\centerline{\bf DIMENSIONAL CROSSOVER AND EFFECTIVE EXPONENTS}
\vskip 24pt
\centerline{Sen-Ben Liao$^{1,2}$\footnote{$^\dagger$}{present address:
Department of Physics, National Chung-Cheng University, Chia-Yi, Taiwan 
R. O. C.; electronic address: senben@phy.ccu.edu.tw}
and Michael Strickland $^1$\footnote{$^\ddagger$}{electronic address:
strickla@phy.duke.edu}}
\vskip 12pt
\centerline{\it Department of Physics $^1$}
\centerline{\it Duke University }
\centerline{\it Durham, North Carolina\ \ 27708\ \ \ U.S.A.}
\vskip 12pt
\centerline{\it and}
\medskip
\centerline{\it Department of Electrophysics $^2$}
\centerline{\it National Chiao-Tung University}
\centerline{\it Hsinchu, Taiwan R. O. C.}
\vskip 1.2in
\vskip 24 pt
\baselineskip 12pt plus 2pt minus 2pt
\centerline{{\bf ABSTRACT}}
\medskip
\medskip

We investigate the critical behavior of the $\lambda\phi^4$ 
theory defined on $S^1\times R^d$ having two finite length
scales $\beta$, the circumference of $S^1$, and $k^{-1}$, the blocking 
scale introduced by the renormalization group transformation.
By numerically solving the coupled differential RG equations for the 
finite-temperature
blocked potential $U_{\beta,k}(\Phi)$ and the wavefunction renormalization
constant ${\cal Z}_{\beta,k}(\Phi)$, we  
demonstrate how the finite-size scaling variable $\bar\beta=\beta k$
determines whether the phase transition is $(d+1)$- or $d$-dimensional  
in the limits $\bar\beta \gg 1$ and $\bar\beta\ll 1$, respectively. 
For the intermediate values of $\bar\beta$, finite-size effects play an
important role. We also discuss the failure of the polynomial 
expansion of the effective potential near criticality.

\vskip 24pt
\vfill
\noindent DUKE-TH-95-101\hfill February 1997
\eject

\centerline{\bf I. INTRODUCTION}
\medskip
\nobreak
\xdef\secsym{1.}\global\meqno = 1
\medskip
\nobreak

The characteristics of a physical system often depend on the energy scale 
of interest 
as well as external environmental variables. Consider QCD as
an example. At distances much less than the
confinement scale $\Lambda_{\rm QCD}^{-1}\sim 1$ fm,
asymptotic freedom gives a vanishing coupling 
strength between the quark and gluon fields.
On the other hand, if the system is coupled to an external heat bath
at temperature $T>\Lambda_{\rm QCD}$, a deconfinement transition takes place,
leading to the formation of a quark-gluon plasma\ref\polonyi. 
Although physics in the low energy or low $T$ limit 
can in principle be accounted for by the fundamental quark 
and gluon degrees of freedom, it is more natural to
use baryons and mesons as the effective degrees of freedom.
This simple illustration shows how effective degrees of 
freedom can be altered by an environmental factor which in this case is $T$. 
The necessity of changing degrees of freedom in order to describe
physical phenomena at different energy scales or environmental conditions 
is often indicative of crossover which is the 
interplay between two types of critical behaviors \ref\lawrie.  

Another class of crossover of considerable
interest is dimensional crossover, in which the system 
undergoes profound changes as its dimensionality, $D$,
is varied.
The dependence of the critical behavior on $D$ is easily understood
from the fact that $D$ and the symmetry group
of the order parameter are the essential 
elements for the classification of universality classes. 
Dimensional crossover can be studied by considering a system having
a finite size in one or more dimensions. Our interest in dimensional
crossover has its origins in the imaginary-time approach to finite-temperature 
field theory.  Within this
formalism the system is defined on the manifold $S^1\times R^d$ with the 
inverse temperature 
$\beta=T^{-1}$ being the circumference of $S^1$, i.e., the system is 
infinite in $d$ dimensions and finite in the remaining one. 
As the imaginary-time ``thickness'' $\beta$ approaches infinity, 
one should recover the expected $T=0$ result which is formally equivalent to 
$R^{d+1}$.
In the high $T$ limit, however,
dimensional reduction takes place, yielding properties which are
characteristic of $R^d$. Thus, decreasing $\beta$ allows 
the system to crossover from $D=d+1$ to $d$. 

The use of finite-size scaling 
arguments \ref\fisher\ to explore dimensional crossover 
has been extensively
discussed by O'Connor, Stephens, van Eijck and collaborators
using the ``Environmentally Friendly Renormalization'' (EFR) 
prescription,
and has had remarkable success in extracting universal quantities such as the
critical exponents and the ratios of critical amplitudes \ref\oconnor;
similar approaches have also been presented in \ref\others.
When the ``environmental'' or ``anisotropic'' factor 
$T$ is such that
$T \gg \mu_{\beta}$, where $\mu_{\beta}$ is the temperature-dependent mass
of the theory, a new type of infrared (IR) divergence independent of 
the ultraviolet (UV) cutoff scale $\Lambda$ appears.
The idea of EFR is to construct 
a set of $T$-dependent counterterms so that not only UV singularities are
subtracted off, but also the IR divergences associated with the ratio
$T/{\mu_{\beta}}$. The difficulty encountered 
in using $T$-independent counterterms
stems from the fact that IR singularities are necessarily 
$T$-dependent. The 
conventional $T$-independent prescription 
fails to describe crossover since the degrees of freedom in 
the crossover regime differ from those near $T\sim 0$. 

While EFR offers valuable insight into dimensional crossover, there exists 
an equally promising non-perturbative tool:  the finite-temperature 
renormalization 
group (RG) \ref\mike\ \ref\sb. 
This RG formulation is based on the Wilson-Kadanoff
blocking transformation \ref\wilson\ and can be illustrated by 
considering the following scalar action
\eqn\slan{ S[\phi]=\int dt\int_{\sx}~\Bigl\{{1\over 2}(\parmu\phi)^2+V(\phi)
\Bigr\},\qquad\qquad   \int_{\sx}=\int d^d{\sx}}
in $d$ spatial plus one time dimensions. 
At finite-temperature the time dimension is compactified
to $S^1$, and one applies the coarse-graining blocking 
transformation to the $d$-dimensional subsystem using
an $O(d)$ symmetric smearing function $\rho_k^{(d)}(\sx)$.
The original field, being periodic in $S^1$, can be expanded as
\eqn\ppx{\phi(\sx,\tau_x)={1\over\beta}
\sum_{n=-\infty}^\infty e^{-i\omega_n\tau_x}\phi_n(\sx)= 
{1\over\beta}\sum_{n=-\infty}^\infty\int_{\s p}
e^{-i(\omega_n\tau_x-\s p\cdot\s x)}\phi_n(\s p), \qquad
\int_{\s p}=\int {d^d\s p\over(2\pi)^d},}
where $\omega_n={2\pi n}/{\beta}$ are the Matsubara frequencies. The
resulting effective blocked field 
is then given by
\eqn\blctk{\phi_k(\s x)={1\over\beta}\int_0^\beta d\tau_y\int_{\s y}
\rho^{(d)}_k(\s x-\s y)\phi(y)={1\over\beta}\int_{|\s p| < k}
e^{i\s p\cdot\s x}\phi_0(\s p)=\Phi(\sx),}
where for simplicity, we have chosen
\eqn\smfrk{\rho_k^{(d)}(\s x)=\int_{|\s p| < k}e^{i\s p\cdot\s x}, }
or $\rho^{(d)}_k(\s p)=\delta_{n,0}\Theta(k-|\s p|)$, using
$\int_0^{\beta}d\tau_x{\rm exp}(-i\omega_n\tau_x)=\beta\delta_{n,0}$. 
This can be
contrasted with $\rho^{(d+1)}_k(p)=\Theta(k-p)$ constructed for 
$R^{d+1}$ which corresponds to the zero-temperature limit of
$S^1\times R^d$ \ref\lp. The advantage of
making the smearing function a sharp cutoff in momentum space is that
the scale $k$ will provide a clear separation between the fast
and slow modes. As can be seen from \blctk, our 
blocking in the imaginary-time formalism leads to an integration over all
modes with $n\ne 0$ as well as the $|\s p|>k$ modes for $n=0$.

In order to probe the physics at an energy scale
$E\sim k$, it is desirable to integrate out the irrelevant microscopic 
degrees of freedom between
$k$ and the UV cutoff $\Lambda$. The naive perturbative
approach in which all the fast-fluctuating modes are integrated out at once
can dramatically change the running parameters due to the  
incorporation of a 
large amount of fluctuations. For example, a significant shift in the mass 
parameter from its bare value $\mu^2_{B}$ to $\tilde\mu^2_{\beta,k}$ is 
generally produced if $\Lambda \gg k$. 
Having large 
dressings means that the running of the parameters of
the theory will not be tracked well by this type of RG trajectory \oconnor;
poor tracking of the running parameters is 
why perturbation theory is plagued by severe IR singularities
in the high-$T$ limit and near
the critical point. 
As an improvement, the Wegner-Houghton differential
RG prescription \ref\wegner\ divides the integration volume into
a large number of thin shells each having a thickness $\Delta k$.
A systematic elimination of each shell is then performed
until the desired scale is reached. In this manner, the continuous 
feedbacks from the higher modes to the lower ones are incorporated.
Moreover, since the running parameters are dressed infinitesimally
along the RG trajectory, the effective degrees of freedom of the theory
are accurately followed. This concept is of key importance in the 
formulation of the exact renormalization group \ref\exact.
In particular, such 
nonperturbative approach would remain applicable in the neighborhood of
crossover where new effective degrees of freedom enter.
Moreover, one can show that the IR divergences are completely lifted as the RG 
improved coupling constant tends to zero at criticality.

The arbitrariness of $k$, the IR cutoff,  
forms the basis for the
momentum RG. By varying $k$ infinitesimally from $k\to k-\Delta k$,
we arrive at the following RG equation for the finite-temperature
blocked potential $U_{\beta,k}(\Phi)$:
\eqn\rgftu{\eqalign{\dot U_{\beta,k}& 
=-{S_dk^d\over\beta}{\rm ln~sinh}\Bigl({\beta\sqrt{k^2
+U''_{\beta,k}/{\calzbk}}\over 2}~\Bigr)\cr
&
=-{S_dk^d\over 2\beta}\Biggl\{\beta\sqrt{k^2+U''_{\beta,k}/{\calzbk}}
+2~{\rm ln}\Bigl[1-e^{-\beta\sqrt{k^2+U''_{\beta,k}/{\calzbk}}}\Bigr]
\Biggr\},}}
where $S_d=2/{(4\pi)^{d/2}\Gamma(d/2)}$ and the dot notation
denotes the operation $k{d/dk}$. However, when no confusion arises, 
the same notation will be used for partial differentiation as well. 
The effect is the wavefunction renormalization constant $\calzbk$ is also
incorporated. In \rgftu, the first and the second terms
may formally be interpreted as the contributions from quantum and
thermal fluctuations, respectively, with the former being
dominant in the low-$T$
regime and the latter in the high-$T$ limit. 
One may regard eq. \rgftu\ as the finite-temperature analog of the 
Wegner-Houghton differential RG equation \wegner. 

From the example of QCD, it is quite clear that
the effective degrees of freedom will vary from one regime to the 
other \ref\paris, and
it would certainly not be suitable to attempt
to use only one particular
set of degrees of freedom to describe physical phenomena at all scales. In
scalar field theory, we shall see that
degrees of freedom characteristic of $(d+1)$ 
dimension can account for the low-$T$ behavior of the system, and at
sufficiently high $T$, the 
corresponding $d$-dimensional counterparts must be used.
In the intermediate range, 
the system exhibits a mixture of the two limiting cases and is 
well described by the RG evolution of $U_{\beta,k}(\Phi)$.
As we shall see later, such dimensional crossover is characterized by
the dimensionless scale $\bar\beta=\beta k$.

In the present work, we follow closely
the formalism developed in \mike\ and continue
to explore the role of 
$U_{\beta,k}(\Phi)$ in describing
physical phenomena at various temperature and momentum scales.
As an extension of our previous work, we couple to the evolution of
$U_{\beta,k}(\Phi)$ 
an additional RG equation arising from the consideration of
$\calz_{\beta,k}(\Phi)$.
Solutions to the two coupled nonlinear partial 
differential equations will provide a smooth connection between the 
small- and large distance physics at arbitrary finite temperature.
In particular, the way in which $\calz_{\beta,k}(\Phi)$ influences
the critical behavior of the system can be seen in a 
rather transparent manner. We shall demonstrate that
our RG approach is equally as ``environmentally friendly'' as that advocated 
in \oconnor.

The organization of the paper is as follows: In Sec. II 
the perturbative one-loop finite-temperature blocked potential 
$\tilde U_{\beta,k}(\Phi)$ and 
the wavefunction renormalization constant $\tilde Z_{\beta,k}(\Phi)$
for one-component scalar theory are derived based on the momentum 
blocking method. In Sec. III we use the results of Sec. II
and generate the coupled RG flow equations 
for the improved $ U_{\beta,k}(\Phi)$ and 
$\calz_{\beta,k}(\Phi)$. In addition to solving these equations
explicitly, we also approximate 
$U_{\beta,k}(\Phi)$ by
polynomial expansion. Sec. IV contains the discussions of
dimensional crossover. It is shown that the phenomenon takes place
approximately at the scale $\bar\beta\sim 1$.
If the system is initially
in a broken phase, at sufficiently high $T$, we expect the symmetry to
be restored. 
Numerical calcualtions of the critical exponents are 
presented in Sec. V for $D$ equal to three and four by
considering the limits $\bar\beta \gg 1$ and $\bar\beta \ll 1$, respectively, 
and are seen to be
in excellent agreement with previous results.
Sec. VI is reserved for summary and discussions.

\medskip
\bigskip
\centerline{\bf II. FINITE-TEMPERATURE SCALAR THEORY}
\medskip
\nobreak
\xdef\secsym{2.}\global\meqno = 1
\medskip
\nobreak

We first consider the following bare action
\eqn\eus{ S[\phi]=\int_0^{\beta}d\tau\int_{\s x}\Biggl\{
{Z\over 2}(\partial_{\tau}\phi)^2+{Z\over 2}
(\nabla\phi)^2+V(\phi)\Biggr\},}
defined on the manifold $S^1\times R^d$, with $Z$ being the bare wavefunction
renormalization constant which conventionally is taken to be unity. 
The action corresponds to
a $(d+1)$-dimensional layered classical system of thickness $\beta$
or a $d$-dimensional
quantum system of ``time'' extent $\beta\hbar$.
Via the coarse-graining blocking transformation, a new blocked action
$\widetilde S_{\beta,k}[\Phi]$ which can be obtained \sb: 
\eqn\cactt{ e^{-\widetilde S_{\beta,k}[\Phi(\sx)]}=\int_{\rm periodic} 
D[\phi]\prod_{\sx}
\delta(\phi_k(\sx)-\Phi(\sx))e^{-S[\phi]},}
where the field average of a given block $\Phi(\sx)$ is chosen to coincide with
the slowly varying background since 
$\phi_k(\sp)=\rho^{(d)}_k(\s p)\phi_n(\s p)
=\Theta(k-|\s p|)\phi_0(\s p)$ with \smfrk. 
An alternative choice for the smearing function is
\eqn\smea{ \rho_{\tilde n, k}^{(d)}(\sx,\tau_x)=\rho_{\tilde n}(\tau_x)
\rho_k^{(d)}(\s x)
={1\over\beta}\sum_{n=-\tilde n}^{\tilde n}\int_{|\s p| < k}
e^{-i(\omega_n\tau_x-\s p\cdot\s x)},}
which gives
\eqn\blc{\eqalign{\phi_k(\sx,\tau_x)&={1\over\beta}\int_0^\beta d\tau_y
\int_{\s y}
\rho^{(d)}_{\tilde n,k}(\s x-\s y,\tau_x-\tau_y)\phi(y)={1\over\beta}
\sum_{n=-\tilde n}^{\tilde n}\int_{|\s p| < k}
e^{-i(\omega_n\tau_x-\s p\cdot\s x)}\phi_n(\s p)\cr
&
={1\over\beta}\int_{|\s p| < k}e^{i\s p\cdot\s x}\Biggl\{
\phi_0(\sp)+e^{-i\omega_1\tau_x}\phi_1(\sp)+e^{i\omega_1\tau_x}\phi_{-1}(\sp)
+\cdots \cr
&\qquad\qquad\qquad
+e^{-i\omega_{\tilde n}\tau_x}\phi_{\tilde n}(\sp)
+e^{i\omega_{\tilde n}\tau_x}\phi_{-\tilde n}(\sp)\Biggr\}=\Phi(\sx,\tau_x).}}
The difference between \smfrk\ and \smea\ is that in the latter, in addition 
to truncating the higher momentum modes by the step function 
$\Theta(k-|\s p|)$, bounds $(\pm\tilde n)$ have been placed 
on the summation over the Matsubara frequencies with $\rho_{\tilde n}(\tau_x)$.
Thus, $\Phi(\sx,\tau_x)$ differs from $\Phi(\sx)$ in that it 
contains contributions from all $\omega_n$ with $n \le \tilde n$.
Nevertheless, since in the high $T$ regime 
all except for the static $n=0$ modes
are strongly suppressed, it suffices to use \blctk\ knowing that
modes with $n\ne 0$ can be treated perturbatively. Notice that the 
form of $\rho_{\tilde n, k}^{(d)}(\sx,\tau_x)$ is not unique; any prescription
that incorporates the $n=0$ mode with $|\sp| < k$ is equally 
suitable. 

The one-loop contribution which takes into
account the quadratic order of the fluctuations can be written as
\eqn\ublok{ \tilde S^{(1)}_{\beta,k}[\Phi]={1\over 2\beta}
\sum_{n=-\infty}^{\infty}\int_{\sx}\int^{'}_{\sp}
{\rm ln}\Bigl[Z\omega_n^2+Z\sp^2+V''(\Phi)\Bigr],\qquad\quad
\int_{\s p}^{'}=S_d\int_k^{\Lambda}dp p^{d-1},}
where $p=|\sp|$ and we have made the substitutions
\eqn\subs{ p_0\longrightarrow\omega_n={{2\pi n}\over\beta},
\qquad\qquad \int_{-\infty}^{\infty}{dp_0\over 2\pi}\longrightarrow
{1\over\beta}\sum_{n=-\infty}^{\infty},}
in going to the imaginary-time formalism. 
We denote quantities computed using the perturbative approach
with a small tilde, to be distinguished from the RG results in the later
sections. At low $T$ where the gap
between the adjacent modes becomes small, the summation over $n$ can be
replaced by an integration and one readily 
recovers the $(d+1)$-dimensional classical theory.  

In the low-energy limit, derivative expansion can be employed. Following
the method by Fraser \ref\fraser, we write the background field
as $\Phi(\sx)=\Phi_0+\tilde\Phi(\sx)$ where $\tilde\Phi(\sx)$ represents
the small spatial inhomogeneities that can be treated perturbatively.
Derivative terms are then produced by treating $\tilde\Phi(\sx)$ and $\sp$
as operators obeying the following commutation relations:
\eqn\commut{ [p_i,\tilde\Phi]=i\partial_i\tilde\Phi, \qquad
[\sp^2,\tilde\Phi]=2ip_i\partial_i\tilde\Phi+\nabla^2\tilde\Phi.}
After carrying out the above procedures, \ublok\ becomes
\eqn\bukk{\eqalign{& \tilde S^{(1)}_{\beta, k}[\Phi]={1\over 2\beta}
\sum_{n=-\infty}^{\infty}\int_x\int^{'}_{\sp}
\Biggl\{ {\rm ln}\bigl[Z\omega_n^2+Z\sp^2+V''(\Phi_0)\bigr]
+{{V'''(\Phi_0)\tilde\Phi+{1\over 2}V''''(\Phi_0)\tilde\Phi^2}\over
{Z\omega_n^2+Z\sp^2+V''(\Phi_0)}} \cr
&
-{1\over 2}\Bigl[{1\over{Z\omega_n^2+Z\sp^2+V''(\Phi_0)}}V'''(\Phi_0)
\tilde\Phi{1\over{Z\omega_n^2+Z\sp^2+V''(\Phi_0)}}V'''(\Phi_0)\tilde
\Phi\Bigr]\Biggr\}+O(\tilde\Phi^3) \cr
&
\equiv\int_x\Biggl\{ -{\tilde Z_{\beta, k}^{(1)}(\Phi_0)
\over 2}
\tilde\Phi\nabla^2\tilde\Phi+\tilde U_{\beta, k}^{(1)}(\Phi_0)
+\tilde U_{\beta, k}^{(1)'}(\Phi_0)
\tilde\Phi+{1\over 2}\tilde U_{\beta, k}^{(1)''}(\Phi_0)\tilde\Phi^2
+\cdots\Biggr\},}}
where $\tilde Z_{\beta, k}^{(1)}(\Phi_0)$ is the one-loop correction to
the wave function renormalization constant.
Upon replacing $\Phi_0$ by $\Phi(\sx)$ 
in \bukk\ \fraser, we are led to
\eqn\fubk{\eqalign{\tilde U^{(1)}_{\beta, k}(\Phi) &={1\over2\beta}\sumn
\int_{\s p}^{'}{\rm ln}\Bigl[Z\omega_n^2+Z\sp^2+V''(\Phi)\Bigr] \cr
&
={1\over 2\beta}\int_{\s p}^{'}\Biggl\{\beta\sqrt{\sp^2+\tilde V''}
+2{\rm ln}\Bigl[1-e^{-\beta\sqrt{\sp^2+\tilde V''}}\Bigr]\Biggr\}+\cdots,}}
and
\eqn\zssy{\eqalign{&\tilde Z^{(1)}_{\beta, k}(\Phi)={({\tilde V}''')^2\over 
{2\beta}}
\sum_{n=-\infty}^{\infty}\int_{\sp}^{'}{{-\sp^2/3+\omega_n^2+\tilde V''}
\over{\bigl(\omega_n^2+\sp^2+\tilde V'')^4}} \cr
&
={({\tilde V}''')^2S_d\over {36\beta}}\int_0^{\infty}ds s^3
\int_k^{\infty}dp p^{d-1}
e^{-(p^2+\tilde V'')s}\sumn\bigl[-p^2+3\omega_n^2+3\tilde V''\bigr]e^{-
\omega_n^2s}.}}
where $\tilde V^{(n)}=V^{(n)}/Z$. 
Performing the summation using the Poisson formulae
%
%
%
%
then gives
\eqn\zst{\eqalign{\tilde Z^{(1)}_{\beta, k}(\Phi)&
={({\tilde V}''')^2S_d\over 72\pi^{1/2}}\int_0^{\infty}ds~s^{3/2}
\int_k^{\infty}dp p^{d-1}e^{-(p^2+\tilde V'')s}\cr
&
\times\Biggl\{{3\over 2}+(-p^2+3\tilde V'')s+2\sumo\Bigl[{3\over 2}
\bigl(1-{\beta^2
n^2\over 2s}\bigr)+(-p^2+3\tilde V'')s\Bigr]e^{-\beta^2n^2/4s}
\Biggr\}.}}
The first two terms inside the braces of the above expression
constitute the usual zero-temperature correction. Eqs. \fubk\ and \zst\
correspond to the standard one-loop 
calculation with derivative expansion and are valid in the
small momentum limit where the effective blocked action 
$\widetilde S_{\beta,k}[\Phi]$ can be characterized by 
$\tilde U_{\beta,k}(\Phi)$,
${\tilde Z}_{\beta,k}(\Phi)$ and other higher-order derivative terms.
However, the presence of IR singularities
in the high $T$ regime will invalidate perturbation theory, and 
the situation can be remedied only if the dominant higher loop corrections
are also resummed. The goal of achieving resummation via RG methods is the
subject of the next section.
%

\medskip
\medskip 

\centerline{\bf III. RENORMALIZATION GROUP IMPROVEMENT}
\medskip
\nobreak
\xdef\secsym{3.}\global\meqno = 1
\medskip
\nobreak
\medskip

Within our framework of finite-temperature RG, there exists two arbitrary
length parameters, $k^{-1}$ and $\beta$ whose ratio defines the dimensionless
quantity $\bar\beta=\beta k$. The momentum scale $k$ naturally 
enters as the effective
IR cutoff for the $R^d$ subsystem during the course of blocking 
transformation. Thus, it is most desirable for
the coarse-graining procedure which employs blocks of size $\beta k^{-d}$
as the effective
degrees of freedom to be continued
until $k^{-1}$, the linear dimension of the blocks becomes comparable to the 
characteristic 
correlation length of the fluctuations in the system, i.e., 
$k^{-1}\sim\xi$. By doing so, the critical behavior of the entire
system can be probed by simply examining the characteristic behavior of any
representative block. 

To set up the RG formalism, we first differentiate \fubk\ with respect to
the arbitrary IR scale $k$:
\eqn\rgfti{\dot{\tilde U}_{\beta, k}
=-{S_dk^d\over{2\beta}}\Biggl\{\beta\sqrt{k^2+\tilde V''}+2{\rm ln}\Bigl[1
-e^{-\beta\sqrt{k^2+\tilde V''}}\Bigr]\Biggr\}.}
%
%
Similarly, after summing over the Matsubara frequencies in \zssy, one
obtains
\eqn\rgzo{\eqalign{\dot{\tilde Z}_{\beta, k}= 
-{({\tilde V}''')^2S_dk^d\over{48\bigl(k^2+\tilde V''\bigr)^{7/2}}}&
\Biggl\{\bigl(-k^2+9\tilde V''\bigr)\Bigl[{1\over 2}+ n_k\bigl(
1+\beta(k^2+\tilde V'')^{1/2}(1+ n_k)\bigr)\Bigr] \cr
&
+ n_k(1+ n_k)\beta^2(k^2+\tilde V'')^2\Bigl[\bigl(
-k^2+3\tilde V''\bigr)(1+2 n_k) \cr
&
-{2\over 3}k^2\beta(k^2+\tilde V'')^{1/2}
(1+6 n_k+6 n^2_k)\Bigr]\Biggr\},}}
where $n_k=\bigl(e^{-\beta\sqrt{k^2+\tilde V''}}-1\bigr)^{-1}$.

The above two differential 
equations are obtained by integrating out each mode independently,
ignoring the feedback from fast modes to slow modes as the IR
cutoff is lowered. Such an ``independent-mode approximation'' (IMA)  
only accounts for contributions up to the one-loop order.
In our RG approach, instead of integrating out all the modes 
between $\Lambda$ and $k$ 
at once, one first divides the integration volume into a large 
number of thin shells of small thickness $\Delta k$, and lower the 
cutoff infinitesimally from $\Lambda\to \Lambda-\Delta k$ until
$\Lambda=k$ is reached. In this manner, we arrive at the 
following coupled RG equations:
\eqn\rgft{\biggl[k\partial_k-{1\over 2}\bigl(d-2+\eta\bigr)\bar\Phi
\partial_{\bar\Phi}+d\biggr]\bar U_{\beta,k}
=-{S_d\over 2}\Biggl\{\bar\beta\sqrt{1+\hat U''_{\beta,k}}+2{\rm ln}
\bigl[1-e^{-\bar\beta\sqrt{1+\hat U''_{\beta,k}} }\bigr]\Biggr\},} 
%
and
\eqn\rgz{\eqalign{& \bigl(k\partial_k-\eta\bigr)\bar{\cal Z}_{\beta, k}\cr
&
= -{({\hat U}'''_{\beta,k})^2{\bar\beta}S_d
\over{48\bigl(1+\hat U''_{\beta,k}\bigr)^{7/2}  }}
\Biggl\{\bigl(-1+9{\hat U}_{\beta, k}''\bigr)
\biggl[{1\over 2}+n_{\bar\beta}\Bigl\{1+\bar\beta\bigl(1+\hat 
U''_{\beta,k})^{1/2}(1+ n_{\bar\beta})\Bigr\}\biggr] \cr
&
+ n_{\bar\beta}(1+n_{\bar\beta})\bar\beta^2(1+\hat U''_{\beta,k})
\Bigl[(-1
+3{\hat U}_{\beta, k}'')(1+2 n_{\bar\beta}) 
-{2\bar\beta\over 3}\bigl(1+\hat U''_{\beta,k})^{1/2}
\bigl(1+6 n_{\bar\beta}+6 n^2_{\bar\beta}\bigr)\Bigr]\Biggr\},}}
where the dimensionless quantities are defined as
\eqn\dimmm{\eqalign{&\bar U_{\beta,k}(\bar\Phi)=\beta k^{-d}U_{\beta,k}
(\Phi), \qquad
\bar\Phi=\beta^{1/2}k^{-(d-2+\eta)/2}\Phi,\qquad \bar\beta=\beta k,\cr
&\bar U^{(m)}_{\beta,k}(\bar\Phi)={{\partial^m \bar U_{\beta,k}(\bar\Phi)}\over
{\partial\bar\Phi^m}}=\beta^{1-m/2}k^{-d+m(d-2+\eta)/2}
U^{(m)}_{\beta,k}(\Phi),\cr
&\bar\mu^2_{\beta,k}=\bar U^{(2)}_{\beta,k}(0)=k^{-2+\eta}\mu^2_{\beta,k},~~
\bar\lambda_{\beta,k}=\bar U^{(4)}_{\beta,k}(0)=\beta^{-1} k^{d-4+2\eta}
\lambda_{\beta,k},\cr
&\bar\calzbk=k^{\eta}\calzbk ,\qquad {\hat U}^{(n)}_{\beta,k}(\bar\Phi)=
{\bar U}^{(n)}_{\beta,k}(\bar\Phi)/{\bar\calzbk(\bar\Phi)}, \cr
& n_{\bar\beta}=\bigl(e^{\bar\beta\sqrt{1+{\hat U}''_{\beta,k}}}
-1\bigr)^{-1}.}}
These two coupled nonlinear partial differential equations provide a 
smooth connection
between the small- and large-distance physics at finite temperature. 
The power of \rgft\ and \rgz\ is that they systematically
incorporate the contributions of a particular mode for the elimination 
of the next, thereby taking into consideration not only the
daisy and superdaisy graphs, but also all higher-order nonoverlapping
loop diagrams.
Notice that in analogy to \rgftu, the first term inside the braces
of \rgft\ 
represents the effect of quantum fluctuations and dominates at low $T$,
and contributions from thermal fluctuations appearing
in the second term become dominant at large $T$.
In Fig. 1 we illustrate the temperature dependence of 
${\cal Z}_{\beta,k=0}(\Phi)$. 
In a similar manner,
the evolution equation for the $m$-point 
scale-dependent vertex function $\bar U^{(m)}_{\beta,k}(\bar\Phi)$
can be written as
\eqn\dirmy{\biggl[k\partial_k-{1\over 2}\bigl(d-2+\eta\bigr)\bigl(m+\bar\Phi
\partial_{\bar\Phi}\bigr)+d\biggr]\bar U^{(m)}_{\beta,k}(\bar\Phi)
=-S_d\partial^m_{\bar\Phi}\Biggl[{\rm ln~sinh}
\Bigl({\bar\beta\sqrt{1+{\hat U}''_{\beta,k}
(\bar\Phi)}\over 2}\Bigr)\Biggr],}
upon differentiating \rgft\ with respect to $\bar\Phi$.

\ifnum\dofig=1
\medskip

\centerline{\epsfbox{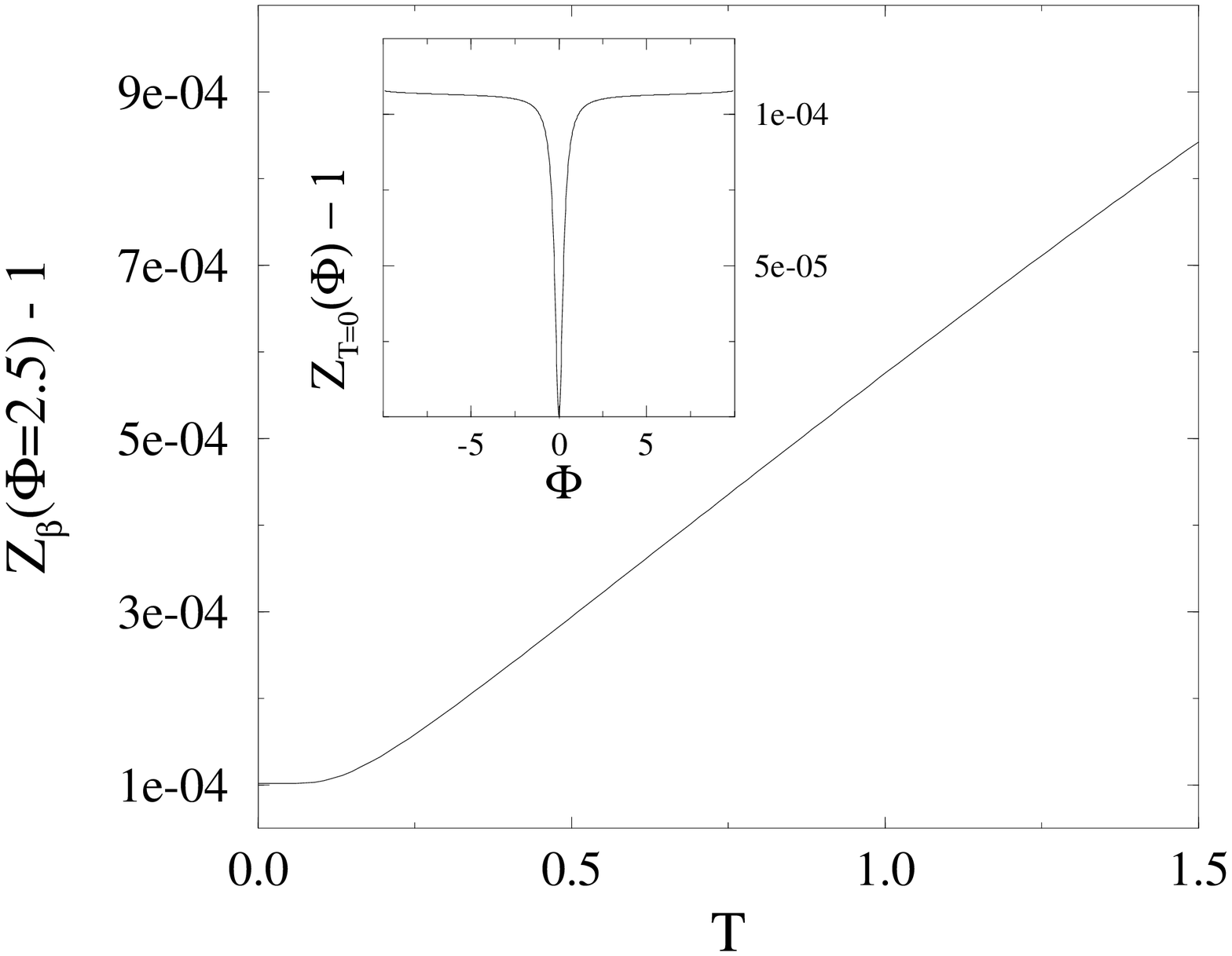}}
\medskip
{\narrower
{\sevenrm
{\baselineskip=8pt
\itemitem{Figure 1.}
Temperature dependence of the wavefunction renormalization constant
$\scriptstyle {\cal Z}_{\beta,k=0}(\Phi)$. Inset shows the 
$\scriptstyle \Phi$ dependence of 
$\scriptstyle {\cal Z}_{\beta,k=0}(\Phi)$ at $\scriptstyle T=0$. 
\bigskip
}}}
\fi

One may compare \rgft\ with the
corresponding zero-temperature $d$-dimensional flow equation for the blocked
potential $\bar U_{k,d}(\bar\Phi_d)$ \lp:
\eqn\foiu{ \biggl[k\partial_k-{1\over 2}\bigl(d-2+\eta\bigr)\bar\Phi_d
\partial_{\bar\Phi_d}+d\biggr] \bar U_{k,d}(\bar\Phi_d)=-{S_d\over 2}
{\rm ln}\Bigl[1+{\hat U}''_{k,d}(\bar\Phi_d)\Bigr],}
%
where 
\eqn\ddimw{\eqalign{&\bar U_{k,d}(\bar\Phi_d)=k^{-d}U_{k,d}(\Phi_d),\qquad
\bar{\cal Z}^{-1}_{k,d}(\bar\Phi_d)=k^{\eta}{\cal Z}^{-1}_{k,d}(\Phi_d),
\qquad \bar\Phi_d=k^{-(d-2+\eta)/2}\Phi_d,\cr
&{\hat U}''_{k,d}=\partial^2_{\bar\Phi_d}{\bar U}_{k,d}/{\bar{\cal Z}_{k,d}},
~~~\bar g^{(m)}_{k,d}=\bar U_{k,d}^{(m)}(0)=k^{-d+m(d-2+\eta)/2} 
g^{(m)}_{k,d},
\qquad \bar\lambda_{k,d}=k^{d-4+2\eta}\lambda_{k,d}.}}
The major difference between these two equations is that the RG flow at finite 
temperature depends on the additional dimensionless parameter 
$\bar\beta$. However, under
extreme temperature conditions, the RG evolution equation can be greatly
simplified. Consider first the high $T$ limit where $\bar\beta \ll 1$. The
main contribution to the flow comes primarily from thermal fluctuations,
the second term in the braces of \rgft. Neglecting 
the effect of quantum fluctuations entirely allows us to write
\eqn\diyd{\eqalign{\biggl[k\partial_k-{1\over 2}\bigl(d-2+\eta\bigr)
\bar\Phi_d\partial_{\bar\Phi_d}+d\biggr]\bar U_{k,d}(\bar\Phi_d)
&=-S_d{\rm ln}\Biggl[{{\bar\beta\sqrt{1+ {\hat U}''_{k,d}(\bar\Phi_d)}}
\over{\bar\beta\sqrt{1+{\hat U}''_{k,d}(0)}}}\Biggr] \cr
&
=-{S_d\over 2}{\rm ln}\Biggl[{{1+{\hat U}''_{k,d}(\bar\Phi_d)}\over{1
+{\hat U}''_{k,d}(0)}}\Biggr],}}
which apart from an irrelevant normalization factor is completely
equivalent to \foiu. The scaling relations between dimensionful and
dimensionless quantities in this region are then given by
\eqn\resca{~~~~~~~~~~~\bar\beta\longrightarrow 
0:\quad \cases{\eqalign{& S^1\times R^d\longrightarrow R^d 
~~~~~~~~~\qquad\bigl(D\longrightarrow d\bigr)\cr
&\bar\Phi\longrightarrow \bar\Phi_d~~~~~~~~~~~~~~~~~~\qquad\bigl(\Phi
\longrightarrow\beta^{-1/2}\Phi_d\bigr) \cr
&\bar U_{\beta,k}\longrightarrow \bar U_{k,d}~~~~~~~~~~~~~\qquad
\bigl(U_{\beta,k}\longrightarrow \beta^{-1} U_{k,d}\bigr)\cr
&\bar{\cal Z}_{\beta,k}\longrightarrow \bar{\cal Z}_{k,d}~~~~~~~~~~~~~\qquad
\bigl({\cal Z}_{\beta,k}\longrightarrow {\cal Z}_{k,d}\bigr) \cr
&\bar g^{(m)}_{\beta,k}\longrightarrow\bar g^{(m)}_{k,d}~~~~~~~~~~~~~\qquad
\bigl(g^{(m)}_{\beta,k}\longrightarrow\beta^{m/2-1}g^{(m)}_{k,d}\bigr). \cr}}}
Thus, we see that in the high-$T$ limit, the system is reduced to
a $d$-dimensional classical theory at zero $T$ \ref\ilawrie. 
On the other hand, in the low $T$ regime where $\bar\beta(1
+{\hat U}''_{\beta,k}(\bar\Phi))^{1/2}\gg 1$, quantum fluctuations prevail
and give
\eqn\cfdsa{ \biggl[k\partial_k-{1\over 2}\bigl(d-2+\eta\bigr)\bar\Phi
\partial_{\bar\Phi}+d\biggr]\bar U_{k,d+1}
=-{S_d\over 2}\sqrt{1+{\hat U}''_{k,d+1}},}
upon neglecting the thermal contributions. 
The $(d+1)$-dimensional characteristic of the system in this limit suggests
the following scaling relations:
\eqn\rescal{\bar\beta\longrightarrow
\infty:\quad \cases{\eqalign{& S^1\times R^d
\longrightarrow R^{d+1}\qquad~~~~~\bigl(D\to d+1\bigr) \cr
&\bar\Phi\longrightarrow \bar\beta^{1/2}\bar\Phi_{d+1}~~~~~~~~~~~~~~ 
\bigl(\Phi\longrightarrow\Phi_{d+1}\bigr) \cr
&\bar U_{\beta,k}\longrightarrow \bar\beta\bar U_{k,d+1}~~~~~~~~~~~~
\bigl(U_{\beta,k}\longrightarrow  U_{k,d+1}\bigr)\cr
&\bar{\cal Z}_{\beta,k}\longrightarrow \bar{\cal Z}_{k,d+1}\qquad~~~~~~~~
\bigl({\cal Z}_{\beta,k}\longrightarrow {\cal Z}_{k,d+1}\bigr) \cr
&\bar g^{(m)}_{\beta,k}\longrightarrow\bar\beta^{1-m/2}\bar g^{(m)}_{k,d+1}
~~~~~~\bigl(g^{(m)}_{\beta,k}\longrightarrow g^{(m)}_{k,d+1}\bigr). \cr}}}
In order to see the equivalence between \cfdsa\ and the RG equation
that possesses $O(d+1)$ symmetry, we make use of the following transformation 
\eqn\cdfs{ S_d\sqrt{1+{\hat U}''_{k,d+1}}
\longrightarrow S_{d+1}{\rm ln}\Bigl[1+{\hat U}''_{k,d+1}\Bigr],}
which is obtained from \mike\
\eqn\csad{\eqalign{&S_d\int_0^kd\sp\sp^{d-1}\sqrt{\sp^2+U''}=
\int^k{d^d\sp\over{(2\pi)^{d/2}}}\sqrt{\sp^2+U''}
=\int^k{d^d\sp\over(2\pi)^{d/2}}\int{dp_0\over 2\pi}
{\rm ln}(p_0^2+\sp^2+U'')\cr
&
={2\Omega_d\over (2\pi)^{d+1}}\int_0^{\pi/2}d\theta{\rm cos}^{d-1}\theta
\int_0^{{\sqrt 2}k}dp~p^d{\rm ln}\bigl(p^2+U''\bigr)
=S_{d+1}\int_0^{{\sqrt 2}k}dp~p^d{\rm ln}\bigl(p^2+U''\bigr),}}
with $\Omega_d=2\pi^{d/2}/{\Gamma(d/2)}$ being the $d$-dimensional solid
angle, and
\eqn\cdfe{\eqalign{ k{d\over dk}&=k\partial_k+k\Bigl(
{\partial\bar\Phi_d\over{\partial k}}\Bigr)\partial_{\bar\Phi_d}
=k\partial_k-{1\over 2}\bigl(d-2+\eta\bigr)\bar\Phi_d
\partial_{\bar\Phi_d} \cr
&
=k\partial_k+k\Bigl({\partial\bar\Phi_{d+1}\over
{\partial k}}\Bigr)\partial_{\bar\Phi_{d+1}}=k\partial_k-{1\over 2}\bigl(d-1
+\eta\bigr)\bar\Phi_{d+1}\partial_{\bar\Phi_{d+1}}.}}
Thus, the $O(d)$-symmetric \cfdsa\ can be equated with \foiu , 
with $d$ replaced by $d+1$. The above scaling patterns illustrate the role
played by the parameter $\bar\beta$ in the crossover from 
$(d+1)-$ to $d$-dimensional and vice versa.
%
Moreover, dimensional crossover
makes no reference to the specific structure of $\bar U_{\beta,k}(\bar\Phi)$;
it only requires that the condition
$\bar\beta(1+\partial^2_{\bar\Phi}\bar U_{\beta,k}(\bar\Phi))^{1/2} \ll 1$
be fulfilled. The manner in which $\bar\beta$ affects $D$ may be 
understood from the following 
heuristic argument: The representative degrees of freedom after applying
the blocking transformation are blocks of finite volume $\beta k^{-d}$ which
becomes infinite as $k\to 0$. The boundary
effect of $S^1$ can be neglected when the linear dimension of the block
falls below $\beta$ and the manifold $S^1\times R^d$ is 
equivalent to $R^{d+1}$ for all practical purpose. 
That is, for $k\to 0$ such that $\bar\beta \gg 1$,
the block volume becomes ``$\infty\times\infty^{d}=\infty^{d+1}$.''
Conversely, when $k^{-1}$ is much larger than $\beta$ such that 
$\bar\beta \ll 1$,
the submanifold $S^1$ becomes ``unnoticed'' and the system undergoes 
a crossover to $R^d$ (block volume $\sim \beta\times\infty^d\sim\infty^d$). 
Thus, the dimensional
crossover scale can be established as $\bar\beta\sim 1$. 

It is also important to point out that the reduction of $D$ 
from $d+1$ to
$d$ in the limit $\bar\beta\to 0$ can be realized in two distinctive physical
situations. The first possibility is to hold $\beta$ fixed while having 
$k\to 0$.
So long as the transition does not take place at 
$T_c=0$, the critical behavior
will always be controlled by the $d$-dimensional fixed point. 
The other possibility, sending $\beta$ to zero while keeping $k$ finite, 
coincides with the well-known
scenario of high-$T$ dimensional reduction. Thus, we see that it is
$\bar\beta$, and not the magnitude of $\beta$ itself, that provides the 
indication of whether the system is in the ``high'' or ``low''
temperature limit. Notice that the latter case
applies irrespective of whether the vacuum symmetry is spontaneously 
broken or not.

\medskip
\medskip
\centerline{\bf B. Polynomial Expansion of $U_{\beta,k}(\Phi)$}
\medskip
\medskip

The nonlinearity of the flow equations \rgft\ and \rgz\ makes analytical 
solutions rather difficult. While numerical solutions can be generated,
very little is known concerning the nature of the operators involved. 
The simplest possible approximation is to write the potential as
\eqn\uexpa{\bar U_{\beta,k}(\bar
\Phi)=\sum_{m=1}^{\infty}{
{\bar g}^{(2m)}_{\beta,k}\over{(2m)!}}\bar\Phi^{2m},\qquad\qquad 
{\bar g}^{(2m)}_{\beta,k}={\bar U}^{(2m)}_{\beta,k}(0),}
where ${\bar g}^{(2)}_{\beta,k}$ and ${\bar g}^{(4)}_{\beta,k}$ can be
identified as the $k$-dependent finite-temperature mass parameter 
$\bar\mu^2_{\beta,k}$ and coupling constant $\bar\lambda_{\beta,k}$, 
respectively. A truncation of the Taylor series expansion
can be made by assuming the smallness of the
higher order terms. Thus, keeping only the leading order
contributions, one could then approximate the theory by the following
coupled flow equations:
\eqn\efmas{ \Bigl[k\partial_k+2-\eta\Bigr]
\bar\mu^2_{\beta,k}=-{{\bar\lbk S_d\bar\beta}\over 4{\bar u_{\bar\beta}}}
{\rm coth}\Bigl({{\bar\beta\bar u_{\bar\beta}}\over 2}\Bigr),}
\eqn\efcou{\eqalign{\Bigl[k\partial_k+\bigl(4-d-2\eta\bigr)\Bigr]
\bar\lambda_{\beta,k}&={3\bar\lbk^2S_d\bar\beta^2\over 8{
\bar u^2_{\bar\beta}}}\biggl\{{1\over{\bar\beta{\bar u}_{\bar\beta}}}
~{\rm coth}
\Bigl({{\bar\beta\bar u_{\bar\beta}}\over 2}\Bigr)
+{1\over 2}{\rm csch}^2\Bigl({{\bar\beta{\bar u}_{\beta}}\over 2}\Bigr)
\biggr\}\cr
&
-{{\bar g^{(6)}_{\beta,k}S_d\bar\beta}\over 4\bar u_{\bar\beta}}
~{\rm coth}\Bigl({{\bar\beta{\bar u}_{\bar\beta}}\over 2}\Bigr),}}
and
\eqn\egssk{\eqalign{ \Bigl[k\partial_k+\bigl(6-2d & -3\eta\bigr)\Bigr]\bar
g^{(6)}_{\beta,k}=-{45\bar\lbk^3S_d\bar\beta\over
16{\bar u}_{\bar\beta}^3}\biggl\{ {\bar\beta^2\over 6}{\rm coth}
\Bigl({{\bar\beta {\bar u}_{\bar\beta}}
\over 2}\Bigr){\rm csch}^2\Bigl({{\bar\beta {\bar u}_{\bar\beta}}
\over 2}\Bigr) \cr
&
+{1\over {\bar u}_{\bar\beta}^2}~{\rm coth}\Bigl({{
\bar\beta {\bar u}_{\bar\beta}}\over 2}\Bigr) 
+{\bar\beta\over 2{\bar u}_{\bar\beta}}{\rm csch}^2\Bigl({{\bar\beta 
{\bar u}_{\bar\beta}}\over 2}
\Bigr)\biggr\} \cr
&+{15\bar\lbk {\bar g}^{(6)}_{\beta,k}S_d\bar\beta\over 
8{\bar u}_{\bar\beta}^2}\biggl\{{1\over {\bar u}_{\bar\beta}}~{\rm coth}
\Bigl({{\bar\beta {\bar u}_{\beta}}\over 2}\Bigr)+{\bar\beta\over 2}
{\rm csch}^2\Bigl({{\bar\beta {\bar u}_{\bar\beta}}\over 2}\Bigr)\biggr\}
+O(\bar g^{(8)}_{\beta,k}),}}
where ${\bar u}_{\bar\beta}=(1+\bar\mu^2_{\beta,k})^{1/2}$. In a 
similar manner, the flow of
$\bar{\cal Z}_{\beta,k}(\bar\Phi)$ can be taken into account by
substituting \uexpa\ into \rgz. 
%
%
In Sec. V we compare the numerical results for the critical exponents 
obtained from keeping the polynomial series up to 
$\bar\Phi^{10}$ with that generated by solving
\rgft\ and \rgz\ directly without resorting to any approximation. 
What we shall find is that the former first converges to the latter, but
then ceases to do so beyond a certain order \ref\morris. The 
nonconverging behavior can be understood from the presence of singularities
in the vertex functions $g^{(2m)}_{\beta,k}$ for $m \ge 4$, as we shall see.

When the symmetry is spontaneously broken
at $T=0$ or when $k$ becomes very large, the potential will have a
nontrivial $k$- and $\beta$-dependent minimum 
$\hat{\bar\Phi}_{\beta,k}$. By assuming the potential to be of the 
form \uexpa,  one my locate $\hat{\bar\Phi}_{\beta,k}$ by solving
\eqn\minp{ 0={{\partial {\bar U}_{\beta,k}}\over{\partial \bar\Phi}}\Big
\vert_{\hat{\bar\Phi}_{\beta,k}}=\sum_{m=1}^{\infty}{{\bar g}^{(2m)}_{\beta,k}
\over{(2m-1)!}}\hat{\bar\Phi}_{\beta,k}^{2m-2}.}
Differentiating \minp\ with respect to ${\rm ln}k$ then yields
\eqn\varik{0=\dot {\bar U}'_{\beta,k}(\hat{\bar\Phi}_{\beta,k})
=\sum_{m=1}^{\infty}{\hat{\bar\Phi}_{\beta,k}^{(2m-3)}\over{(2m-1)!}}
\biggl\{\dot {\bar g}^{(2m)}_{\beta,k}
\hat{\bar\Phi}_{\beta,k}+(2m-2){\bar g}^{(2m)}_{\beta,k}
\dot{\hat{\bar\Phi}}_{\beta,k}\biggr\},}
or
\eqn\varri{ \dot{\hat{\bar\Phi}}_{\beta,k}=- {{\sum_{m=1}^{\infty}{{\dot 
{\bar g}^{(2m)}_{\beta,k}}\over{(2m-1)!}}\hat{\bar\Phi}_{\beta,k}^{2m-2}}\over
{{\sum_{m=1}^{\infty}{{{\bar g}^{(2m)}_{\beta,k}}\over{(2m-2)!}}
\hat{\bar\Phi}_{\beta,k}^{2m-3}}}}=-{{\dot {\bar\mu^2}_{\beta,k}
+{\dot{\bar\lambda}_{\beta,k}
\over 3!}\hat{\bar\Phi}_{\beta,k}^2+{\dot{\bar g}^{(6)}_{\beta,k}\over 5!}
\hat{\bar\Phi}_{\beta,k}^4+\cdots}\over{{2{\bar\lambda}_{\beta,k}\over 3!}
\hat{\bar\Phi}_{\beta,k}+ {4{\bar g}^{(6)}_{\beta,k}\over 5!}
\hat{\bar\Phi}_{\beta,k}^3+ {6{\bar g}^{(8)}_{\beta,k}\over 7!}
\hat{\bar\Phi}_{\beta,k}^5+\cdots}},}
which allows us to explore the variation of $\hat{\bar\Phi}_{\beta,k}$ 
with $k$. While the flow of the theory can be monitored near the origin 
$\bar\Phi=0$ by defining the vertex functions there, one may also
track the evolution of the theory defined at the scale-dependent minimum, i.e.,
instead of \uexpa, we expand the blocked potential as
\eqn\uexpan{ \bar U_{\beta,k}(\bar\Phi)=\sum_{m=0}^{\infty}
{\hat{\bar g}^{(m)}_{\beta,k}\over {m!}}
\bigl(\bar\Phi-\hat{\bar\Phi}_{\beta,k}\bigr)^m,\qquad 
\hat {\bar g}^{(m)}_{\beta,k}
=\bar U^{(m)}_{\beta,k}(\hat{\bar\Phi}_{\beta,k}).}
By solving the RG equations explicitly, we illustrate in Fig. 2 the 
pattern of symmetry restoration above $T_c$.

\ifnum\dofig=1
\medskip

\centerline{\epsfbox{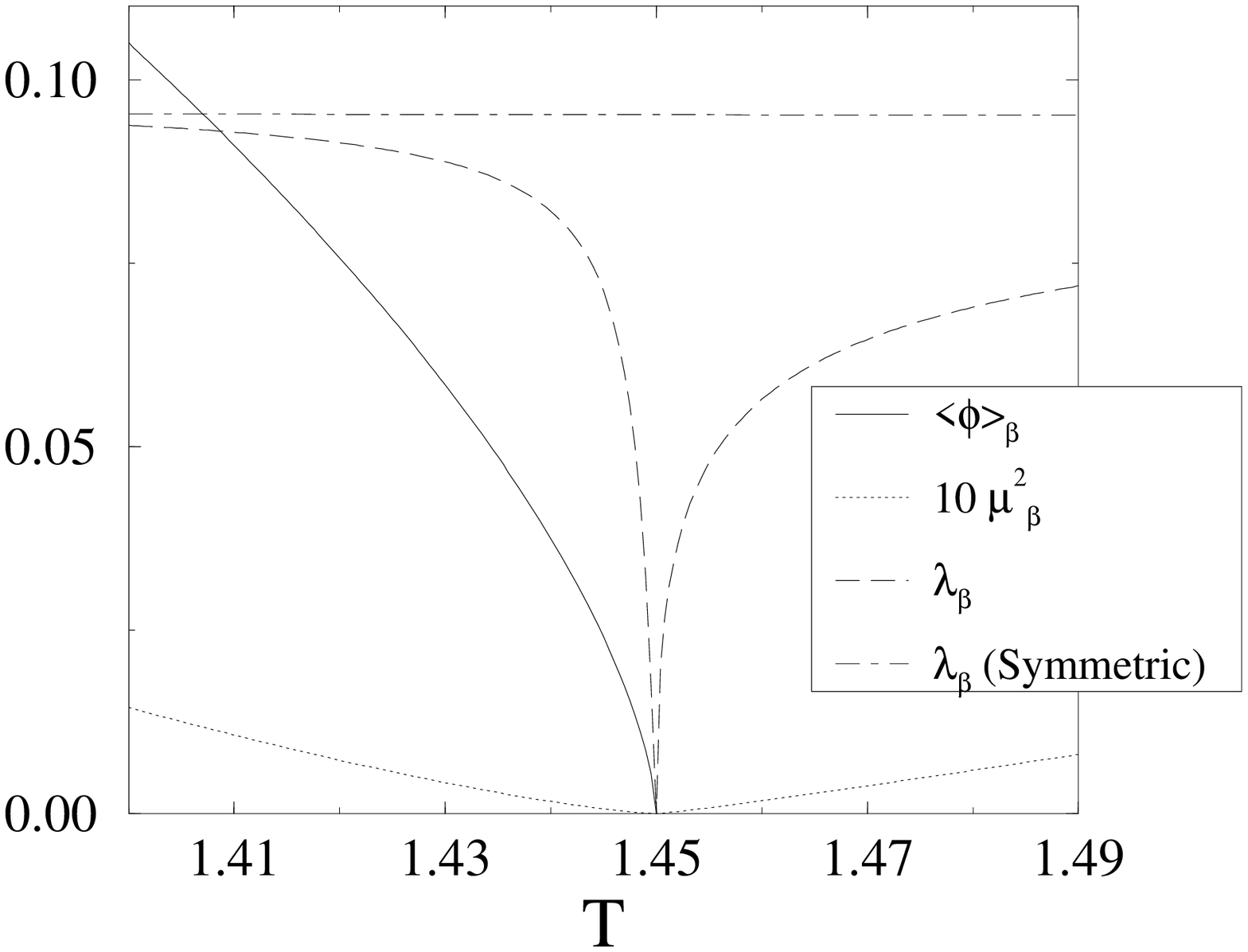}}
\medskip
{\narrower
{\sevenrm
{\baselineskip=8pt
\centerline{
Figure 2.  Blocked potential $\scriptstyle U_{\beta,k}(\Phi)$ near 
$\scriptstyle T_c$. The transition is second order.
}
\bigskip
}}}
\fi

\medskip
\medskip

\centerline{\bf IV. DIMENSIONAL CROSSOVER AND THE CRITICAL EXPONENTS}
\medskip
\nobreak
\xdef\secsym{4.}\global\meqno = 1
\medskip
\nobreak
\centerline{\bf A. Dimensional Crossover}
\medskip
\medskip

In exploring the critical behavior of our blocked system, with
$k^{-1}$ being the characteristic linear dimension 
for the $R^d$ submanifold, it is desirable to have the
coarse-graining procedure
repeated until the block length becomes comparable to the correlation
length $\xi$ which conventionally is measured by the inverse
of the effective mass parameter, i.e., $\xi\sim\mu^{-1}_{\beta}$.
With $k\sim \mu_{\beta}$, as $\mu_{\beta}\to 0$
in the neighborhood of the transition where the volume of individual 
blocks tend to infinity ($k=0$), the correlation length also becomes infinite. 
As we have seen before, whether the transition is $d+1$- or $d$-dimensional 
is intimately related to the value of $\bar\beta$.

In the Introduction, we also argued that beyond the crossover regime, the 
set of effective degrees of freedom is expected to change.
To identify the new degrees of freedom associated with small $\bar\beta$ 
limit, one first makes use of \ppx\ to decompose the original unblocked 
field $\phi(\sx)$ into the static ``light'' mode $\phi_0(\sx)$ and
the ``heavy'' mode $\phi_{n\ne 0}(\sx)$. In this manner, the classical 
action in \eus\ can be
rewritten as \ref\rivers 
\eqn\uss{ S_{\beta}[\phi]=\beta\int_{\sx}\Bigl[{Z\over 2}
(\nabla\phi_0)^2+V(\phi_0)\Bigr]+\int_0^{\beta}d\tau_x\int_{\sx}
\Bigl\{{Z\over 2}\sum_n^{'}\phi_n\bigl(\omega_n^2-\nabla^2\bigr)
\phi_{-n}+\Delta V(\phi_0,\phi_n)\Bigr\},}
where the prime notation implies that the summation is over $n\ne 0$ only.
For the familiar $\lambda\phi^4$ theory considered here, 
\uss\ becomes
\eqn\usu{\eqalign{ &S_{\beta}[\phi]=\beta\int_{\sx}\Biggl\{ \Bigl[{Z\over 2}
(\nabla\phi_0)^2+{\mu^2\over 2}\phi_0^2+{\lambda\over 4!}\phi_0^4\Bigr]
+{1\over 2}\sum_n^{'}\phi_n\Bigl(Z\omega_n^2-Z\nabla^2+\mu^2
+{\lambda\over 2}\phi_0^2\Bigr)
\phi_{-n} \cr
&
+{\lambda\over 6}\phi_0\sum_{n_1,n_2,n_3}^{'}\phi_{n_1}\phi_{n_2}\phi_{n_3}
\delta(n_1+n_2+n_3)+{\lambda\over 4!}\sum_{n_1,\cdots, n_4}^{'}\phi_{n_1}
\cdots\phi_{n_4}\delta(n_1+n_2+n_3+n_4)\Biggr\}.}}
Clearly in the limit $\beta\to 0$, the heavy modes are
strongly damped in the Boltzmann sum and become decoupled, leaving the
static light mode $\phi_0(\sx)$ as the new effective degree of freedom
for the reduced $d$-dimensional system described by
$S_{k,d}[\phi_0]=\beta^{-1}S_{\beta}[\phi]$ \ref\ginsparg. Even though such
decoupling strictly speaking holds only up to $O(\bar\mu^2_{\beta})$ 
\ref\landsman, the dimensionally reduced prescription has been demonstrated
to be a good approximation in the small $\bar\mu_{\beta}$ limit \mike.
While contributions from the $n\ne 0$ Matsubara modes can be
handled perturbatively, the static processes described by the $n=0$ sector
must be treated by non-perturbative techniques.
This justifies our use of the smearing function $\smfrk$ which defines the
static effective blocked fields $\Phi(\sx)$ shown in \blctk.
%
%

In the extreme limits $\bar\beta \gg 1$ and
$\bar\beta \ll 1$ the critical behavior of the system exhibit
$(d+1)-$ and $d$-dimensional characteristics, respectively. What about the 
intermediate values of $\bar\beta$? Do they correspond to any
{\it physical} non-integer dimensionality? The answer is negative since
in principle one cannot associate $S^1\times R^d$ with some 
effective manifold $R^{d_{\rm eff}}$. What we shall find is that in
the intermediate range, finite-size effect can change
the values of the exponents significantly.

\medskip
\medskip
\centerline{\bf B. Phase Transition and Effective Critical Exponents}
\medskip
\medskip

For definiteness we take $d=3$ or $S^1\times R^3$.
We start off by making the thickness of the bulk system infinite, i.e., 
$\beta=\infty$ and gradually decrease its value. In addition, we 
assume that the system is initially in 
the broken phase with a negative 
renormalized parameter $\tilde\mu^2_R < 0$ at $T=0$. As $T$ is raised, 
in the limit $k\to 0$
one expects a second-order phase transition at $T_c$ 
where the symmetry of the system is restored. 

The volume of a real system, however, can never be infinite;
the Monte Carlo RG simulations, too, are performed on finite lattices. 
When dealing with such finite systems, one often 
employs the finite-size scaling assumption to extract the information
associated with the infinite-volume limit. It turns out that 
with our RG prescription the validity
of the finite-size scaling hypothesis can readily be tested 
by focusing on the behavior of each individual block of volume $\beta k^{-d}$.
To accomplish this, we note that although phase transitions 
in a strict sense  can occur only in the infinite volume limit, 
a $k$-dependent ``pseudo-transition'' temperature $T_c(k)$ can be 
defined as the scale at which the minimum as well as the mass 
parameter vanish continuously {\it within} each block, i.e.,
\eqn\frty{ U'_{\beta_c(k),k}(0)=0,\qquad {\rm and}
\qquad U''_{\beta_c(k),k}(0)=0.}
Our $T_c(k)$ is analogous to the rounding temperature commonly encountered
in the Monte Carlo RG simulations. One may justify the use of \frty\ 
as the working definition of the pseudo-transition by noting that in
principle there always exists a $T_c(k)$ at which
the effective mass parameter $U''_{\beta_c(k),k}$ vanishes 
regardless of the size of the block, provided that
$T$ be kept substantially below $\Lambda$ in order
to be considered as physical. While the finiteness of the blocks makes 
the transition temperature necessarily $k$-dependent, the true 
critical temperature is given by $T_c=T_c(k=0)$ which is independent of $k$. 
We emphasize that our system is always
infinitely large with an infinite number of blocks which represent the 
effective  degrees of freedom. However, what we are primarily interested 
here is to elucidate
the critical behavior of the whole system by looking into just one single
block whose size varies between zero ($\beta\Lambda^{-3}$) and infinity. 
The measured critical exponents will exhibit
three- and four-dimensonal characteristics for $T_c(k=0)$ and
$T_c(k=\Lambda)$, respectively, with the latter being the original 
``unblocked'', bare system. 

There exists two separate methods for investigating the
``critical'' behavior of a given block. The first one is to fix the theory
at $T_c(k)$ and examine the variation of the thermodynamical
quantities with $T$ in this vicinity.
In the second approach, one first determine the true $T_c$ by taking the
$k\to 0$ limit, and then inquire how the thermodynamical quantities 
deviate with $k$ with the help of finite-size scaling assumption. The  
schematic diagram of the two methods are given in Fig. 3 below. 

\ifnum\dofig=1
\medskip

\centerline{\epsfbox{fig10.eps}}
\medskip
{\narrower
{\sevenrm
{\baselineskip=8pt
\itemitem{Figure 3.}
Schematic diagram of the two methods used for measuring the critical exponents.
The left block summarizes method (1) and the right method (2).
\bigskip
}}}
\fi

\medskip
\medskip
\par\hang\noindent{(1)} pseudo-transition at $T_c(k)$
\medskip
\medskip

If we stop the RG flow at some arbitrary scale $k$, then in the neighborhood
of $T_c(k)$, the {\it effective}
critical exponents can be related to the thermodynamical
quantities in the following manner:
\eqn\crits{\eqalign{
&\chi^{-1}=\mu^2_{\beta,k}\sim |T-T_c(k)|^{\gamma_{\rm eff}},
\qquad\qquad\qquad~~~~ \beta\to\beta_c(k), \cr
&\eta_{\rm eff}=-{{\partial~{\rm ln}{\cal Z}_{\beta_c,k}
(\hat\Phi_{\beta_c,k})}\over{\partial {\rm ln}k}}, 
\qquad\qquad\qquad\qquad~~ \hat\Phi_{\beta_c,k}\to 0, \cr
&\hat\Phi_{\beta,k}\sim |T-T_c(k)|^{\beta_{\rm eff}},
\qquad\qquad\qquad\qquad~~~~~~~ \beta\to\beta_c(k), \cr
&\hat\Phi_{\beta_c(k),k}\sim h^{1/{\delta_{\rm eff}}},
\qquad\qquad\qquad\qquad\qquad~~~~~~~~h\to 0.}}
Notice that while the susceptibility diverges at $T_c(k)$, the correlation
length given by $\xi\sim (\mu^2_{\beta_c(k),k} + k^2)^{-1/2}\sim 
k^{-1}$ remains finite as long as $k\ne 0$. This also implies that
the conventional expression
$\xi\sim |T-T_c(k)|^{-\nu_{\rm eff}}$ is not directly applicable when 
dealing with a finite system whose correlation length cannot diverge.
In the above, the
exponent $\delta_{\rm eff}$ gives a measure of how the ``magnetization''
$\hat\Phi_{\beta,k}$ varies with an external field $h$ as $h\to 0$ at
$T_c(k)$. Its measurement is achieved by coupling to $\Phi$
a constant source term $h$.

The effective exponents obtained in this manner will certainly depend on 
$\bar\beta$,
e.g., $\gamma_{\rm eff}=\gamma_{\rm eff}(\bar\beta)$, and the dependence is
lifted only when $\bar\beta=0$ or $\bar\beta=\infty$. In fact, we shall see 
that $\gamma_{\rm eff}(0)=\gamma_3$ and $\gamma_{\rm eff}(\infty)
=\gamma_4$, where the subscripts correspond to the physical dimensionality
$D$. To what extent the system will deviate from
its true critical behavior can be analyzed from the relative 
shift of the pseudo-critical temperature due
to the finiteness of the block:
\eqn\tshif{ \lim_{k\to 0} |T_c-T_c(k)| \sim k^{\theta}.}
Numerically, we find $\theta=1.49 \pm .01$ using the polynomial expansion of
$U_{\beta,k}(\Phi)$ up to $O(\Phi^8)$. The value of $\theta$ in general
can be related to $\nu_{\rm eff}$ as $\theta = 
\nu^{-1}_{\rm eff}(0)=\nu^{-1}_3$. Notice that one may also take the opposite 
limit $k\to\Lambda$. In this case, $\theta\approx 2$, leading to
$\nu_4=1/2$. 

\medskip
\medskip
\par\hang\noindent{(2)} finite-size scaling:
\medskip
\medskip

Based on the principle of finite-size scaling, one may assume   
that near $T_c$ thermodynamical quantities depend only on the dimensionless
ratio $\ell/{\xi}$ where $\ell$ and $\xi$ are, respectively, 
the characteritic finite size and the correlation
length of the system. For example, the
susceptibility $\chi$ can be written as
\eqn\fssc{ \chi=\xi^{\gamma_{\rm eff}/{\nu_{\rm eff}}}f\bigl(
{\ell\over\xi}\bigr),}
where $f(r)$ is a scaling function with the property $f(r)\to {\rm const.}$
as $r\to\infty$. With the presence of two finite length scales, 
$\ell=\beta$ and $k^{-1}$, in our system, two different scalings are
expected. For $\ell=k^{-1}$, using the relation
$\xi^{1/{\nu_{\rm eff}}}\sim |T-T_c(k)|^{-1}$, \fssc\ can then be 
rewritten as 
\eqn\fsf{ \chi=|T-T_c(k)|^{-\gamma_{\rm eff}}f\bigl({|T
-T_c(k)|^{-\nu_{\rm eff}}\over k}\bigr),}
which shows that $f$ depends on the scaling variable
$Y_{\beta} =|T-T_c(k)|/k^{1/{\nu_{\rm eff}}}$.
This ``$Y_{\beta}$'' scaling
prescription which corresponds to the pseudo-transition we have just
addressed, explores the $T$ dependence near $T_c(k)$
while keeping $k$ fixed. 
The pattern of crossover associated with $Y_{\beta}$ is
depicted in Fig. 4.

\medskip
\bigskip
\ifnum\dofig=1
\medskip

\centerline{\epsfbox{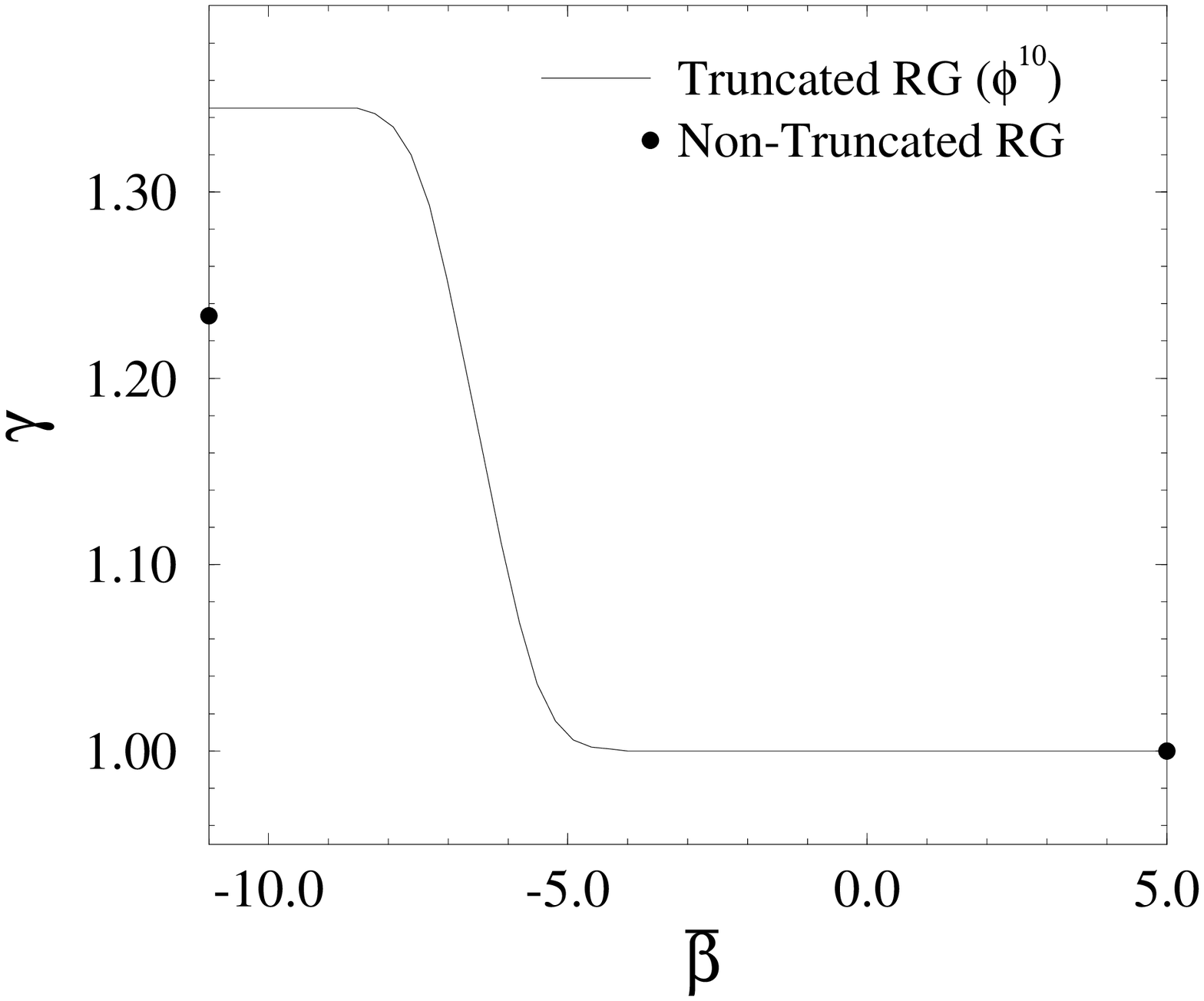}}
\medskip
{\narrower
{\sevenrm
{\baselineskip=8pt
\itemitem{Figure 4.}
Effective critical exponent 
$\scriptstyle \gamma_{\rm eff}$
as function of (a) $\scriptstyle {\rm log}_{10}(T - T_c(k))$  and (b) 
$\scriptstyle {\rm log}_{10}((T - T_c(k))/k^{1/\nu_{\rm eff}})$, 
where the three-dimensional
value $\scriptstyle \nu_3^{-1}=\nu^{-1}_{\rm eff}(0)=1.49 \pm 0.01$ 
is obtained via measuring the shift $\scriptstyle |T_c-T_c(k)|$ 
due to the finiteness of $\scriptstyle k$.
The different symbols correspond to different values of $\scriptstyle k$.
\bigskip
}}}
\fi

On the other hand, if we choose $\ell=\beta_c$, the $k$-independent inverse
critical temperature and let    
the correlation length of the finite system be given by 
the maximum linear dimension of the block, i.e., $\xi\sim k^{-1}$, 
\fssc\ then takes on the form
\eqn\nffs{ \chi=k^{-\gamma_{\rm eff}/{\nu_{\rm eff}}}{\tilde f}\bigl(
\beta_ck\bigr)=k^{-\gamma_{\rm eff}/{\nu_{\rm eff}}}{\tilde f}\bigl(
\bar\beta_c\bigr),}
where $\tilde f$ is another scaling function with $Y_k=\bar\beta_c$ being
the scaling variable. This is completely consistent with our claim that
$\bar\beta$ sets the scale of dimensional crossover.
Thus, instead of extracting the exponents from the $T$
dependence of the thermal parameters near $T_c(k)$, we may  
go directly to the true $T_c$ and 
explore the $k$ dependence of thermodynamical quantities, viz,
\eqn\fss{\chi^{-1}\sim  
k^{\gamma_{\rm eff}/{\nu_{\rm eff}}}, \qquad\quad
\hat\Phi_{\beta_c,k}\sim k^{\beta_{\rm eff}/{\nu_{\rm eff}}}.}
The pattern of crossover associated with $Y_k$ is exemplified in Fig. 5.

\ifnum\dofig=1
\medskip

\centerline{\epsfbox{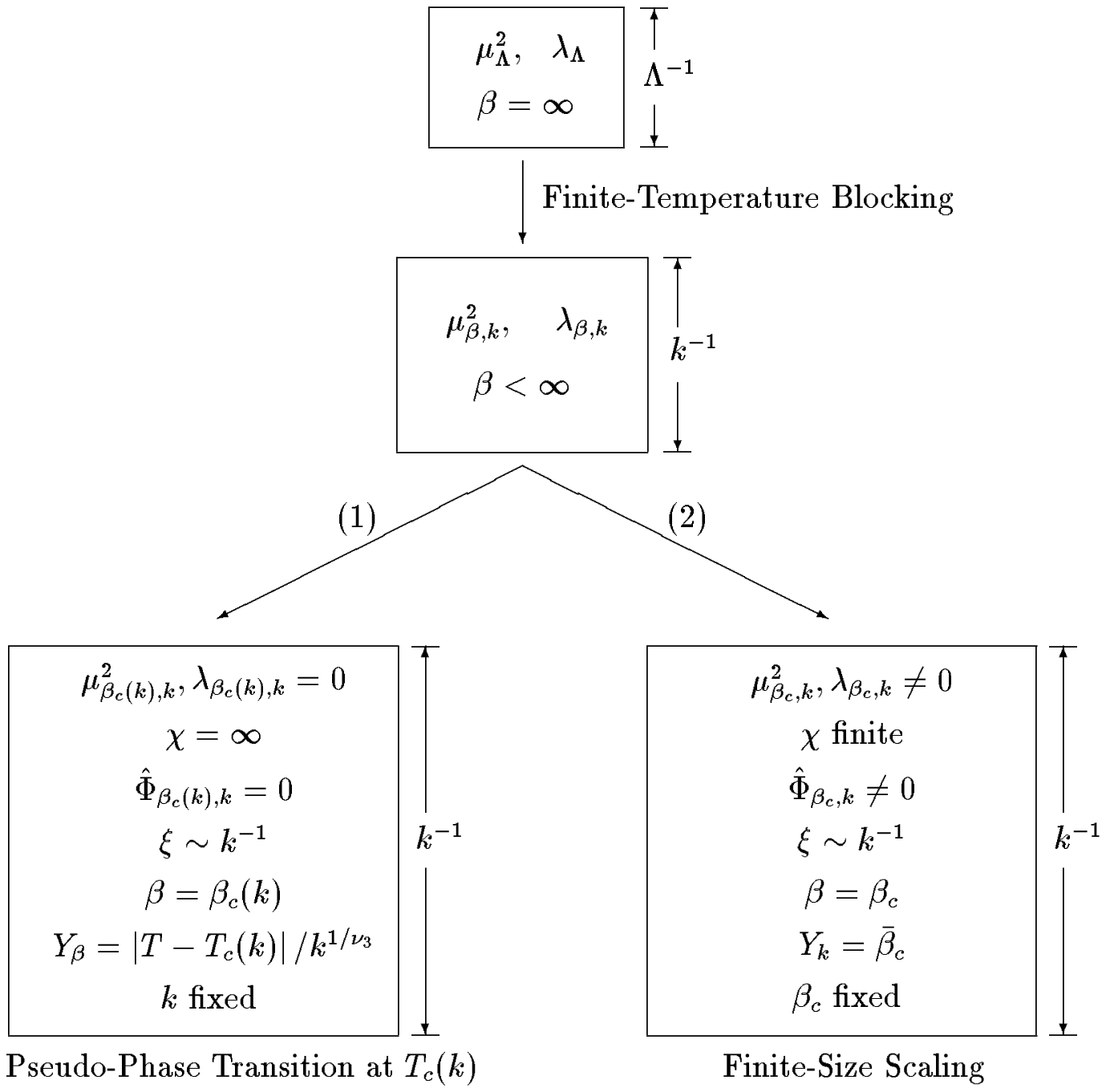}}
\medskip
{\narrower
{\sevenrm
{\baselineskip=8pt
\itemitem{Figure 5.}
Dependence of $\scriptstyle \gamma_{\rm eff}$ on 
$\scriptstyle \bar\beta_c$ obtained via polynomial expansion of 
$\scriptstyle U_{\beta,k}(\Phi)$ up to
$\scriptstyle O(\Phi^8)$. The values of $\scriptstyle \gamma_3$ and 
$\scriptstyle \gamma_4$ calculated from 
integrating the dimensionful version of (3.3) and (3.4) are also included.
\bigskip
}}}
\fi

It is important to note that although thermodynamical 
quantities may diverge or vanish at $T_c(k)$, they remain finite at 
the true critical temperature $T_c$ for $k \neq 0$, as can be seen
from Fig. 6 for the minimum $\hat\Phi_{\beta,k}$. 
Comparing the two procedures, we see that
while examining the $T$ dependence of the physical quantities 
near $T_c(k)$ allows us to extract the critical exponents,  
exploring the $k$ dependence
at $T_c$ in the second approach yields exponents as ratios of 
$\nu_{\rm eff}$ which must be extracted from
\eqn\nuef{ \lim_{k\to 0}~\lim_{T\to T_c} |T^2-T^2_c(k)|\sim 
k^{1/{\nu_{\rm eff}(k)}}.}
Comparison with \tshif\ implies $\theta=\nu^{-1}_{\rm eff}(0)$.
Various ratios of the exponents extracted 
by polynomial truncation without the inclusion
of $\calzbk$ are summarized in Table 2. In addition, 
since $\lambda_{\beta,k}/{\bar\mu_{\beta,k}}$ 
approaches a constant as $k\to 0$ and $T\to T_c$ and 
$\lambda_{\beta_c,k=0}=0$ \mike\ \ref\wetterich, we see that both 
$\lambda_{\beta,k}$ and $\mu_{\beta,k}$ must vanish at the same rate, i.e.,
\eqn\laj{ \lambda_{\beta,k=0} \sim |T-T_c|^{\zeta_{\rm eff}(0)} 
\sim |T-T_c|^{\nu_{\rm eff}(0)},}
which allows for the identification $\zeta_{\rm eff}(0) =\nu_{\rm eff}(0)$ 
when $\eta_{\rm eff}=0$. In other words, neglecting the effect of
$\calzbk$, $\nu_{\rm eff}(0)$ can be deduced by examining how
$\lambda_{\beta}$ vanishes as $T_c$ is approached.

\ifnum\dofig=1
\medskip

\centerline{\epsfbox{fig9.eps}}
\medskip
{\narrower
{\sevenrm
{\baselineskip=8pt
\itemitem{Figure 6.}
Dependence of the minimum $\hat\Phi_{\beta,k}$, on the
IR cutoff $\scriptstyle k$ at the critical temperature
$\scriptstyle T=T_c(k=0)$. Notice that $\scriptstyle\hat\Phi_{\beta,k}$ 
vanishes only when $\scriptstyle k = 0$.
\bigskip
}}}
\fi

\medskip
\medskip
\centerline{\bf V. NUMERICAL METHODS AND RESULTS}
\medskip
\nobreak
\xdef\secsym{5.}\global\meqno = 1
\medskip
\medskip

We now describe the details pf how 
the differential RG equations \rgft\ and \rgz\ are solved numerically.
The evolution begins at
the UV cutoff scale $\Lambda$ where all thermal effects vanish and the 
running parameters take on the bare values which we denote with the
subscript $B$.  We choose $\mu^2_B < 0$ and a
small positive value for the coupling constant $\lambda_B$. 

\ifnum\dofig=1
\medskip

\centerline{\epsfbox{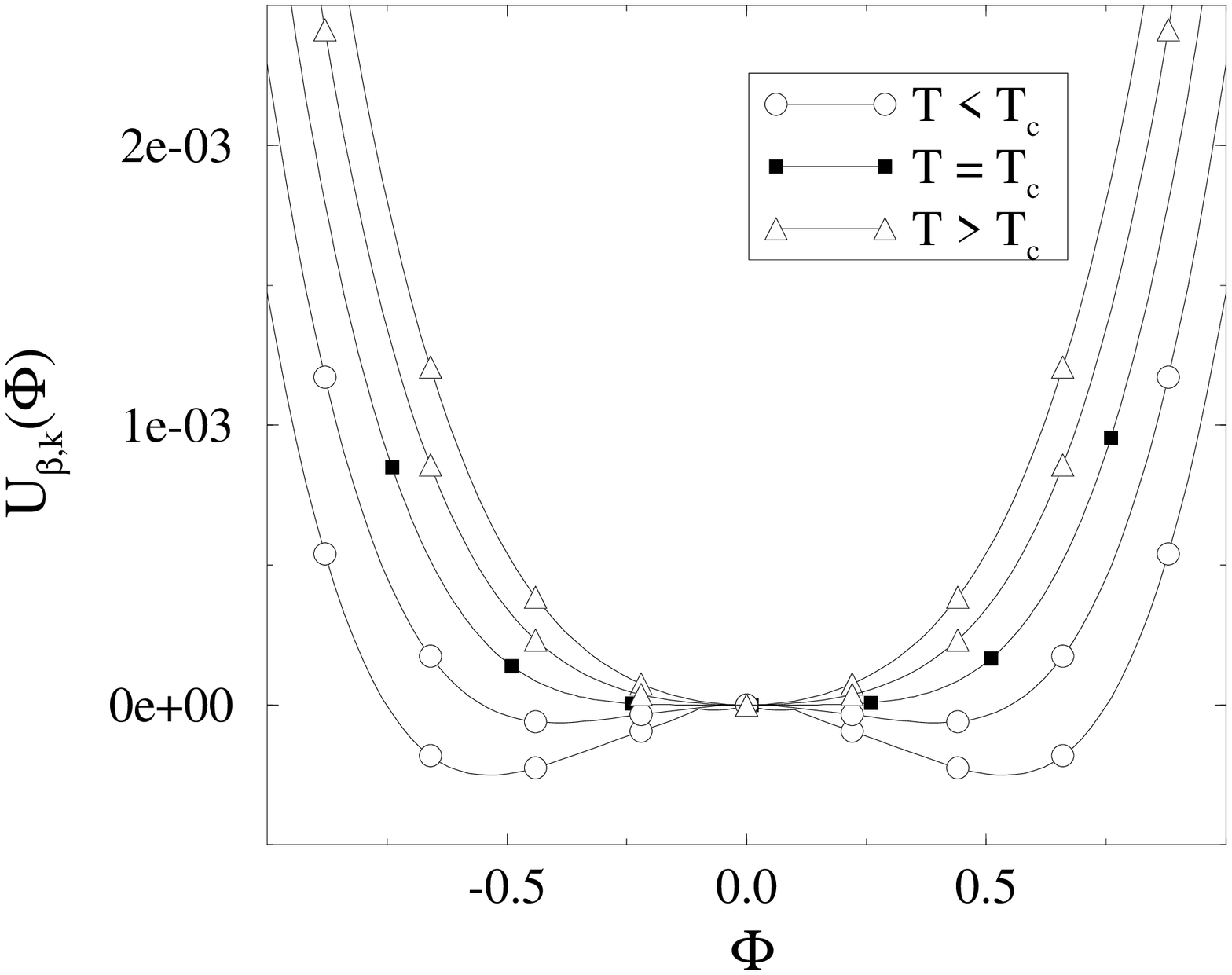}}
\medskip
{\narrower
{\sevenrm
{\baselineskip=8pt
\itemitem{Figure 7.}
Temperature dependence of the thermal mass parameter
$\scriptstyle \mu^2_{\beta}$, coupling constant 
$\scriptstyle \lambda_{\beta}$, and 
the expectation value $\scriptstyle <\phi>_{\beta}$ near 
$\scriptstyle T=T_c$. The $\scriptstyle T$ 
dependence of $\scriptstyle \lambda_{\beta}$ in the symmetric phase is also 
illustrated.
\bigskip
}}}
\fi
\goodbreak
The symmetric phase is characterized by a
positive value of the renormalized mass $\tilde\mu^2_R > 0$ even if the
initial value $\mu^2_B < 0$. On the other hand, if $\tilde\mu^2_R < 0$ 
the theory is in the spontaneously broken phase. 
In Fig. 7 we depict the dependence of the minimum, 
mass, and coupling constant near $T_c(k=0)$.

As noted before, phase transitions strictly speaking can only take place at
$k=0$ where the block volume becomes infinite; however, to map out the 
entire crossover regime, one must study the dependence of 
$U_{\beta,k}(\Phi)$ on $\bar\beta_c$. 
To accomplish this we integrate the {\it dimensionful} versions of
\rgft\ and \rgz\ as well as the truncated flow equations such as
\efmas\ - \egssk\ 
down to a small value $k \neq 0$
for which the pseudo-transition temperature is $T_c(k)$.  The value 
of $k$
can then be adjusted in order to probe different values of $\bar\beta_c$.
Decreasing $k$ corresponds to enlarging the size of the blocks.
The gradual change of the effective exponents with $\bar\beta$
is illustrated in Fig. 4 where the
different symbols correspond to different values of $k$. 

From Fig. 4 (a), we see that true scaling takes place only near integer
dimensions where $k$ is either very large or vanishing. In these
neighborhoods, the measured values 
of the effective exponents are approximately constant for a large range of 
$T-T_c(k)$ with only minute deviations. However, in
the intermediate regime, complete
scaling does not occur and the effective exponents can only be 
approximated by fixing the value of $k$ and selecting
a narrow range of $|T-T_c(k)|$. On the other hand, when
parameterized by the finite-size scaling variable $Y_{\beta}$,
the data collapse nicely
into one curve of $\gamma_{\rm eff}$ which smoothly interpolates between
$\gamma_4$ and $\gamma_3$, as can be seen from Fig. 4(b).

The numerical results indicate that the smaller the value of $\bar\beta$,
the closer the system is to the three-dimensional critical behavior.
On the other hand, a large value of $\bar\beta$ implies
smaller blocks and the exponents will crossover to
the four-dimensional values. This can be explained by observing that
as the individual blocks shrink, the system begins to flow back to
the original four-dimensional unblocked system and the bare theory is
recovered when the linear dimension of the block becomes $\Lambda^{-1}$.
Since thermal fluctuations are highly suppressed at $k\sim \Lambda$,
quantum fluctuations become dominant and ``recreate'' the extra
dimension, making the effective dimensionality four.. 

The truncated flow equations were integrated using a fifth-order
adaptive step-size Runge-Kutta integrator allowing for precise
determination of the thermal parameters near $T_c(k)$.  
In Fig. 5 the $\bar\beta$
dependence of the critical exponent $\gamma_{\rm eff}$ calculated from
an eighth order polynomial truncation of our flow equations is illustrated.
The exponents are independent of the values of
the bare mass parameter, and coupling constant, as required by
universality. By comparing the results obtained with and without truncation,
we see that the latter yields a better agreement with the experimental values.
While the crossover scale is expected to be $Y_k^{1/{\nu_{\rm eff}}}=
\bar\beta_c=1$, the truncated RG equation shows that the
three-dimensional characteristics of the theory is uncovered only when 
$\bar\beta_c$ is substantially less than unity ($\approx 10^{-6}$ from
the Fig. 5). Increasing the order of truncation only affects the result
slightly. Although qualitatively correct, this may 
be an indication of the inadequacy of the polynomial 
approximation in exploring the nonperturbative phenomenon of dimensional
crossover. 
%

In both Figs. 4 and 5 the values of the critical exponents
calculated by explicitly integrating over \rgft\ and \rgz\ without 
polynomial truncation are also given. In carrying out the calculation,
$U_{\beta,k}(\Phi)$ was discretized
with $\Delta\Phi = 0.01$ and the cutoff was lowered systematically from
$k=\Lambda$ to $k=0$ using finite differences.  The $k$ steps were 
second-order Runge-Kutta and the $\Phi$ derivatives were calculated
using a five-point difference.  
Unfortunately, due to the amount of computer time required for
making each run, we were unable to produce plots of the effective
exponents as a function of the continuous variable $\bar\beta_c$ 
from the nontruncated RG equations. Instead,
the dependence of the effective exponents on $\bar\beta$
which takes on continuous values is reported by using the truncated flow 
equations obtained in Sec. III. 


We list in Table 1 the values of the critical exponents 
obtained from various orders of polynomial truncation of \uexpa\ along 
with those calculated by solving \rgft\ and \rgz\ directly without 
truncation. As mentioned before, the three- and the four-dimensional limits
are reached by dialing the value of $\bar\beta$ to be such that
$\bar\beta \ll 1$ and $\bar\beta \gg 1$, respectively. 
>From the Table, one
sees that ${\cal Z}_{\beta,k}$ only gives a minute corrections to
the exponents. Nevertheless, its consideration is crucial for extracting
the anomalous dimension $\eta_3$. 
The results agree very well with 
that derived from $\epsilon$ expansion up to $\epsilon^5$
and five loops \ref\fiveloop. Note that presently we were only able to extract
$\nu_{\rm eff}$ indirectly from studying the $T$ dependence of 
$\lambda_{\beta,k=0}$ in \laj\ using truncated polynomial expansion. 
The measurement of $\alpha_{\rm eff}$ turns out to be rather difficult.
We also comment that extracting the critical
exponents $\beta_{\rm eff}$, $\eta_{\rm eff}$, and $\delta_{\rm eff}$ 
using the truncated flow
equations requires an expansion about the $k$-dependent minimum 
$\hat\Phi_{\beta,k}$ of the blocked potential $U_{\beta,k}(\Phi)$. 
However, we find that the polynomial expansion 
around $\hat\Phi_{\beta,k}$ given by \uexpan\ fails near $T_c$ even 
though away from $T_c$ it agrees well with that obtained using \uexpa.
This is due to the nonanalytic behavior of
$U_{\beta_c,k}(\Phi)$ as $k\to 0$.

\ifnum\dofig=1
\bigskip
\centerline{\sevenrm \vbox{\tabskip=0pt \offinterlineskip
\def\tablerule{\noalign{\hrule}}
\halign to435pt{\strut#&\vrule#\tabskip=1em plus2em&
  #\hfil& \vrule#& #\hfil& \vrule#&
  #\hfil& \vrule#& #\hfil& \vrule#&
  #\hfil& \vrule#& #\hfil& \vrule#\tabskip=0pt\cr\tablerule
&&\multispan{11}\hfil {\bf Measured Critical Exponents for $\scriptstyle 
D = 4$} \hfil&\cr\tablerule
&& \omit\hidewidth  $\scriptstyle O(\Phi^{(2m)})$ 	\hidewidth
&& \omit\hidewidth $\scriptstyle \gamma$	\hidewidth
&& \omit\hidewidth $\scriptstyle \zeta$ ($\nu$)	\hidewidth
&& \omit\hidewidth $\scriptstyle \beta$	\hidewidth
&& \omit\hidewidth $\scriptstyle \eta$	\hidewidth
&& \omit\hidewidth $\scriptstyle \delta$	\hidewidth	&\cr\tablerule
&& NT	    && 1.0000 && 0.5000 && 0.5000 && 0.0000 && 3.000 &\cr
&&          &&        &&        &&        &&        &&       &\cr\tablerule
&&\multispan{11}\hfil {\bf Measured Critical Exponents for $\scriptstyle 
D = 3$  } \hfil&\cr\tablerule
&&   2	  && 1.054 $\pm$ 0.002 	&& 0.527 $\pm$ 0.003 &&         &&  &&        &\cr\tablerule
&&   3			&& 1.171 $\pm$ 0.004	&& 0.585 $\pm$ 0.003 &&         &&  &&        &\cr\tablerule
&&   4			&& 1.345 $\pm$ 0.004	&& 0.672 $\pm$ 0.004 &&         &&  &&        &\cr\tablerule
&&   5			&& 1.504 $\pm$ 0.004	&& 0.744 $\pm$ 0.004 &&         &&  &&        &\cr\tablerule
&&   NT		        && 1.234  $\pm$ 0.01	&&        && 0.315 $\pm$ 0.007 && 0.0000 && 4.65 $\pm$ 0.1 &\cr\tablerule
&&   NT + $\scriptstyle\calzbk$ && 1.241  $\pm$ 0.008	&&  && 0.321 $\pm$ 0.008 && 0.036 $\pm$ 0.005  && 4.63 $\pm$ 0.1 &\cr
&&  &&  &&  &&  &&  &&
&\cr\tablerule
&&\multispan{11}\hfil {\bf Best Calculations to Date$^*$}
\hfil&\cr\tablerule
&& $\scriptstyle \epsilon^5$&& 1.2390 $\pm$ 0.0025 && 0.6310 $\pm$ 0.0015 && 0.3270 $\pm$ 0.0025 && 0.0375 $\pm$ 0.0025 &&  4.814 $\pm$ 0.015 &\cr
&& &&  &&  &&  &&  &&
&\cr\tablerule
&&\multispan{11}\hfil {\bf Experimental Values$^{**}$ } \hfil&\cr\tablerule
&& &&  1.23-1.25 && 0.624-0.626 && 0.316-0.327 && 0.016-0.06 && 4.6-4.9 &\cr\tablerule
}}}
\noindent$^*$From Zinn-Justin \ref\zinn. 
$^{**}$From Goldenfeld \ref\goldenfeld.
\bigskip
{\narrower
{\sevenrm
{\baselineskip=7pt
\itemitem{Table 1.}
Critical exponents as function of the level of truncation along with the best 
calculations to date and experimental values.  NT indicates results obtained
from the nontruncated RG equations \rgft\ and \rgz.
\bigskip
}}}
\fi

In Table 2 we show the $k$ dependence of the $(2m)$-point vertex functions 
$g^{(2m)}_{\beta,k}$ up to $m=5$. From the Table, 
we see that 
the vertex functions diverge in the limit $k\to 0$ for $m \ge 4$, and the
diverging behaviors $(\sim k^{3-m})$ are in 
accord with \dimmm\ where 
one writes $g^{(2m)}_{\beta,k}=\beta^{-1+m}k^{3-m(1+\eta{\rm eff})}\bar 
g^{(2m)}_{\beta,k}$ for $d=3$, apart from $\eta_{\rm eff}$. 
According to the Table, 
$g^{(6)}_{\beta_c,k}$ approaches a constant as $k\to 0$; however,
by taking into consideration the anomalous dimension, one would find that
it diverges as $k^{-3\eta_{\rm eff}}$. 
Notice that the values $a_2=1.998\pm 0.005$ 
and $a_4=0.997\pm 0.004$ measure,
respectively, the ratios of the effective critical exponents 
$\gamma_{\rm eff}/{\nu_{\rm eff}}$ and $\zeta_{\rm eff}/{\nu_{\rm eff}}$.
The fact that $a_4=\zeta_{\rm eff}/{\nu_{\rm eff}}$ is close to unity
gives support to our previous claim that 
$\nu_{\rm eff}$ can be approximated by $\zeta_{\rm eff}$. We also see that
the manner in which the vacuum expectation value $\hat\Phi_{\beta_c,k}$
vanishes as function of $k$ allows for the identification 
${\beta_{\rm eff}/{\nu_{\rm eff}}}=0.51\pm 0.01$. After extracting
$\zeta_{\rm eff}=\nu_{\rm eff}$ by measuring the $k$ dependence of 
$|T-T_c(k)|$ as $k \to 0$ using \nuef, all other
exponents can be deduced from finite-size data. The procedure is 
analogous to
the standard approach employed in the Monte Carlo simulations of
finite-size systems. Using $\nu^{-1}_{\rm eff}(0)=\nu_3^{-1}=1.49 \pm 0.01$, 
we see that the exponents extracted from Table 2 are in good agreement
with those measured by examining the dependence
of the thermal parameters near $T_c(k)$. 

\ifnum\dofig=1
\bigskip
\centerline{\vbox{\tabskip=0pt \offinterlineskip
\def\tablerule{\noalign{\hrule}}
\halign to255pt{\strut#&\vrule#\tabskip=1em plus2em&
  #\hfil& \vrule#& #\hfil& \vrule#\tabskip=0pt\cr\tablerule
&& \omit\hidewidth  $g_{\beta_c,k}^{(2m)}\sim k^{a_{2m}}$ 	\hidewidth
&& \omit\hidewidth $a_{2m}$	\hidewidth	&\cr\tablerule
&&   $g^{(2)}_{\beta_c,k}$	&& 1.998 $\pm 0.005
=\gamma_{\rm eff}/{\nu_{\rm eff}}$ &\cr\tablerule
&&   $g^{(4)}_{\beta_c,k}$	&& 0.997 $\pm 0.004 
=\zeta_{\rm eff}/{\nu_{\rm eff}}$	&\cr\tablerule
&&   $g^{(6)}_{\beta_c,k}$	&& 0.002 $\pm$ 0.004 	&\cr\tablerule
&&   $g^{(8)}_{\beta_c,k}$	&& -1.001 $\pm$ 0.005 	&\cr\tablerule
&&   $g^{(10)}_{\beta_c,k}$	&& -2.002 $\pm$ 0.005 	&\cr\tablerule
&&   $\hat\Phi_{\beta_,k}$   && 0.51 $\pm 0.01 
=\beta_{\rm eff}/{\nu_{\rm eff}}$ &\cr\tablerule
}}}
\bigskip
{\narrower
{\sevenrm
{\baselineskip=7pt
\centerline{
Table 2. $\scriptstyle k$ dependence of vertex functions at 
$\scriptstyle T_c$ in the absence of wavefunction renormalization.
}}}}
\bigskip
\fi

The numerical values we obtained using polynomial truncation in 
$U_{\beta,k}(\Phi)$ are in excellent agreement with \ref\dattanasio\ and
\morris. 
However, as demonstrated by Morris who incorporated polynomial contributions
up to $O(\Phi^{50})$ \morris, the critical exponents derived in this 
manner do not uniformly converge to the experimental values but 
instead exhibit oscillatory behavior. Although the
oscillation has not yet been observed in our truncated
scheme which includes contributions only up to $\Phi^{10}$, we remark that
this generic feature is another indication of the
breakdown of the polynomial expansion of $U_{\beta,k}(\Phi)$ in \uexpa.

The presence of the nonanalytic structure in the blocked potential
can be seen readily by noting that in the zero-temperature
``dressed'' $(V''_R(\Phi)\to V''_k(\Phi))$ approach, the potential
contains a term of the 
form $(k^2+V_k''(\Phi))^2{\rm ln}(1+V_k''(\Phi)/k^2)$, where $V_k''(\Phi)
=\mu_k^2+\lambda_k\Phi^2/2$ \lp. When $V_k''(\Phi)/k^2$ is small, the 
logarithmic term may be expanded as ${\rm ln}(1+{V_k''(\Phi)/k^2})
=\sum_{m=1}^{\infty}(-1)^{(m+1)}(V_k''(\Phi)/k^2)^m/m$ which allows us
to parameterize the blocked potential as polynomials of 
$\Phi$. While such expansion 
is justified in the large $k$ limit, it generally breaks down in the
IR limit where $k\to 0$ even though $\lambda_k$ may be small. In particular,
when $V''_k$ vanishes slower than $k^2$, the logarithmic dependence
of $\Phi$ can no longer be expanded as power series. If one insists 
on making the expansion,
the resulting series will be divergent with alternating sign in $\lambda_k$.  
Similar arguments also hold for finite-temperature systems. 
We emphasize that the IR singularities encountered in
$g^{(2m)}_{\beta,k}$ for $m \ge 4$ illustrated in Table 2 
are merely an artifact which
arises from the breakdown of polynomial expansion of $U_{\beta,k}(\Phi)$
close to $T_c$ and $k=0$ where both the mass
parameter $\mu^2_{\beta_c,k=0}$ and the coupling strength 
$\lambda_{\beta_c,k=0}$ vanish. 
Thus, we conclude that both the presence of
IR divergences and oscillation of the critical exponents around experimental
values are all due to the breakdown of polynomial
expansion close to $T_c$ in the IR regime. 

We find further evidence for this claim by examining the $\Phi$ 
dependence of the derivatives of the blocked potential close to
$T_c$.  In Fig.89, we plot the fourth derivative of the
blocked potential, $U^{(4)}_{\beta_c,k}(\Phi)$, solving the
nontruncated RG equations \rgft\ and \rgz.
Had a fourth order truncation of the blocked potential
been sufficient, i.e., $U_{\beta_c,k}(\Phi)=\mu^2_{\beta_c,k}\Phi^2/2+
\lambda_{\beta_c,k}\Phi^4/4!+O(\Phi^6)$, one then will have
$U^{(4)}_{\beta_c,k}(\Phi)=\lambda_{\beta_c,k}$ which for any given $k$ is 
a constant independent of $\Phi$. While the prediction made by polynomial
truncation is reasonable for large $k$, it is clearly unreliable
in the small $k$ limit. As depicted in Fig. 8, in the IR regime
$U^{(4)}_{\beta_c,k}(\Phi)$ is a complicated nonanalytic function
of $\Phi$ and not a constant! It turns out that a large number of
terms are required if one wishes to 
characterize the $\Phi$ dependence of $U^{(4)}_{\beta,k}(\Phi)$ by
powers series.
For example, the inclusion of the sixth-order term would 
yield a parabolic shape to the
$U^{(4)}_{\beta_c,k}(\Phi)$ near $\Phi=0$ and other higher-order contributions
with increasing magnitude and oscillating signs must be included to
fit the plateau at large $\Phi$ \oconnor. That is, a simple
polynomial fit seems feasible for sufficiently small $\Phi$, but 
becomes unnatural in the plateau region. The difficulty
encountered in fitting $U^{(4)}_{\beta_c,k}(\Phi)$ with power series of $\Phi$ 
casts doubts on the use of polynomial expansion of $U_{\beta,k}(\Phi)$
in the limit $k->0$.

\ifnum\dofig=1
\medskip

\centerline{\epsfbox{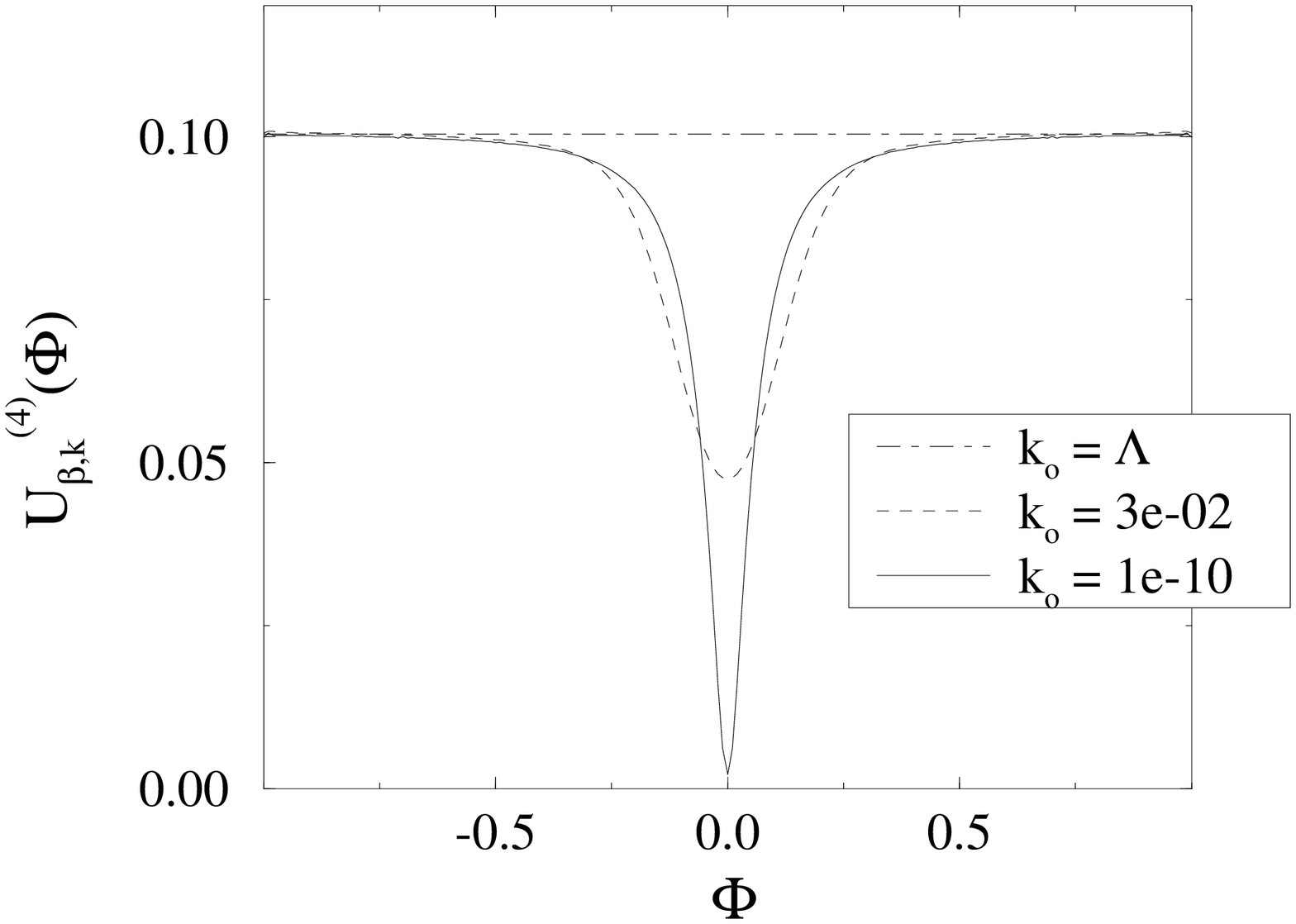}}
\medskip
{\narrower
{\sevenrm
{\baselineskip=8pt
\itemitem{Figure 9.}
$\scriptstyle U^{(4)}_{\beta,k}$ as a function of $\scriptstyle \Phi$ close to 
$\scriptstyle T_c$ 
($\scriptstyle |T - T_c(k)|\sim 10^{-5}$) calculated with 
the nontruncated RG equations. The UV cutoff is taken to be 
$\scriptstyle\Lambda=20$.
\bigskip
}}}
\fi

\medskip
\medskip
\centerline{\bf VI. SUMMARY AND DISCUSSIONS}
\medskip
\nobreak
\xdef\secsym{6.}\global\meqno = 1
\medskip
\nobreak

We have demonstrated in this paper the applicability of RG in describing
dimensional crossover. The nonperturbative nature
of the RG formulation allows us to establish a smooth connection
between the physics in $d+1$ and $d$ dimensions. The 
scale at which crossover takes place is shown to be at 
$\bar\beta\sim 1$. 
Through numerical integration of the coupled RG equations in \rgft\ and
\rgz, the dependence of the effective critical exponents on $\bar\beta$ was
explored. We find that in the limits $\bar\beta \gg 1$ and 
$\bar\beta \ll 1$, the transitions 
can be associated with $(d+1)$ and $d$ dimensions, respectively.
In the intermediate values of $\bar\beta$, finite-size effects are 
important. 

The effective exponents were extracted using two different methods:
In the first approach, we measure the effective critical exponents
by studying the temperature dependence of the 
thermodynamical quantities near $T_c(k)$ with the true critical 
behavior obtained when $k\to 0$. 
In the second approach, we fixed the temperature at $T_c$ and 
examined the $k$ dependence of the thermal parameters.  
Invoking finite-size scaling 
arguments, the critical exponents were obtained as ratios of 
$\nu_{\rm eff}(k)$ 
which was determined using the relation $\lim_{k\to 0}
|T^2-T^2_c(k)|\sim k^{1/{\nu_{\rm eff}(k)}}$ given in \nuef.
Our measurements of the effective critical exponents for
$D=4$ ($k\to\Lambda$) and $D=3 ~(k\to 0)$
are in excellent
agreement with the established results, and are more conclusive
than those presented in \wetterich\ where
a smooth regulator was used. 
In the intermediate range of $\bar\beta$ where scaling is only
approximate, the exponents are deduced by fixing the value of $k$
and restricting the range of $|T-T_c(k)|$. 
The numerical results derived from both
schemes are completely consistent and allow for a smooth interpolation of
the effective exponents between the three- and 
four-dimensional values. 

The success of our 
formalism in exploring the physics of crossover is attributed to its 
accurate tracking of the effective
degrees of freedom at any arbitrary point in the two-parameter space
spanned by the flow parameters $\beta$ and $k$. 
Our approach 
allows us to resum a large class of diagrams that are beyond the 
five-loop calculations up to $\epsilon^5$ carried out in 
\fiveloop. In addition, it is computationally far more advantageous
to solve the differential RG equations \rgft\ and \rgz\ than
to work out hundreds of Feynman graphs to achieve the same level of
accuracy. While our RG equations take into consideration all possible
nonoverlapping Feynman graphs to infinite loop orders, the $\epsilon$
expansion technique provides individual treatments to distinct 
Feynman graphs, both overlapping and nonoverlapping, only to a finite
loop order. It remains an interesting issue to explore the role of
the overlapping graphs that are left out in our RG prescription.
Nevertheless, we believe that our formalism is superior to the $\epsilon$
expansion since the critical exponents deduced from the former are
manifestly finite and agree remarkably well with experiments;
as for the latter, it is in reality an asymptotic series in which
truncations must be made in order to attain agreement with the experimental
measurements. 

In light of the success of our RG prescription in the investigations of 
dimensional crossover as well as critical behavior of the scalar
$\lambda\phi^4$ model, one can readily extend the formalism
to $O(N)$ symmetry in order to study the XY ($N=2$) model, Heisenberg 
($N=3$) model, and the large $N$ limit \ref\largen. Systems at
lower dimensionality can also be studied since the validity and generality
of our treatment can be tested by comparing the results with the
exact solutions possible in certain two-dimensional systems. For the XY
model, spontaneous symmetry breaking can take place for $D > 2$ at
finite $T_c$ but at exactly $D=2$ the system exhibits the well known
Kosterlitz-Thouless phase transition driven by vortices
which are
the topologically singular spin configurations. 
It would be interesting to see if our formalism can capture the
topological crossover as the effective dimension changes.
Another possible application is the consideration of the $3+1$
$U(1)$ scalar QED which has same critical behavior as the three-dimensional
superconductor. Since the effects of quantum fluctuations are
also incorporated in a systematic manner, one may also consider
quantum critical phenomena as in \ref\suranyi.
The RG formalism is not just limited to scalar theories, but
can be implemented in fermionic systems in a similar fashion, as has been
carried out recently in \ref\comellas, and in gauge theories \ref\gauge\
\ref\wkrg.
These issues shall be addressed in the future publications.

\goodbreak
\bigskip
\centerline{\bf ACKNOWLEDGEMENTS}
\medskip
\nobreak
We thank J. Adler, R. Brown, S. Huang, S. Matinyan,
B. M\"uller and J. Socolar for 
stimulating discussions and helpful comments on the manuscript. S.B.L is
grateful to Prof. T.-W. Chiu of the Department of Physics at 
the National Taiwan University for the
hospitality and fruitful discussion during his visit.  M.S. would like
to thank D. O'Connor for email discussion.
This work was
supported in part by the U. S. Department of Energy (Grant No. 
DE-FG05-90ER40592), the North Carolina Supercomputing
Center, and the National Science Council of Taiwan, R. O. C.

\medskip
\medskip
\centerline{\bf REFERENCES}
\medskip
\medskip
\nobreak
\item{\polonyi} see for example, J. Polonyi, ``Physics of the
Quark-Gluon Plasma,''hep-ph/9509334; and B. M\"uller, 
{\it The Physics of the Quark-Gluon Plasma} (Springer-Verlag, 1985).
\medskip
\item{\lawrie} see for example, I. D. Lawrie, {\it J. Phys.}
{\bf A9} (1976) 961, {\it J. Phys.} {\bf C11} (1978) 1123 and 3857;
D. Amit and Y. Goldschmidt, {\it Ann. Phys.} {\bf 114} (1978) 356.
\medskip 
\item{\fisher} M. E. Fisher and M. N. Barber, {\it Phys. Rev.
Lett.} {\bf 28} 1516.
\medskip
\item{\oconnor} D. O'Connor and C. R. Stephens, {\it Int. J.
Mod. Phys.} {\bf A9} (1994) 2805 and {\it Nuc. Phys.} {\bf B360} (1991)
297 and {\it J. Phys.} {\bf A25} (1992) 101; 
F. Freire and C. R. Stephens, {\it Z. Phys.}
{\bf C60} (1993) 127; M. A. van Eijck {\it et al}, {\it Int. J. Mod. Phys.} 
{\bf A10} (1995) 3343.
\medskip
\item{\others} D. Schmeltzer, {\it Phys. Rev.} {\bf B32} (1985) 7512;
E. Frey, {\it Physica} {\bf A221} (1995) 52.
\medskip 
\item{\mike} S.-B. Liao and M. Strickland, {\it Phys. Rev.}
{\bf D52} (1995) 3653.
\medskip 
\item{\sb} S.-B. Liao and J. Polonyi, {\it Nucl. Phys.}
{\bf A570} (1994) 203c; S.-B. Liao, J. Polonyi and D. P. Xu, {\it Phys. Rev.}
{\bf D51} (1995) 748.
\medskip
\item{\wilson} K. Wilson, {\it Phys. Rev.} {\bf B4} (1971) 3174;
K. Wilson and J. Kogut, {\it Phys. Rep.} {\bf 12C} (1975) 75.
\medskip
\item{\lp} S.-B. Liao and J. Polonyi, {\it Ann. Phys.} {\bf 222} (1993) 122
and {\it Phys. Rev.} {\bf D51} (1995) 4474.
\medskip
\item{\wegner} F.J. Wegner and A. Houghton, {\it Phys. Rev} 
{\bf A8} (1972) 401.
\medskip
\item{\exact} 
J. F. Nicoll, T. S. Chang and H. E. Stanley, {\it Phys. Rev.
Lett.} {\bf 32} (1974) 1446, {\bf 33} (1974) 540; {\it Phys. Rev.}
{\bf A13} (1976) 1251;
J. Polchinski, {\it Nucl. Phys.} {\bf B231} (1984) 269;
A. Hasenfratz and P. Hasenfratz, {\it ibid.} {\bf B270} (1986) 687;
C. Wetterich, {\it ibid.} {\bf B352} (1990) 529; 
N. Tetradis and C. Wetterich, {\it ibid.} {\bf B398} (1993) 659;
P. Hasenfratz and J. Nager, {\it Z. Phys.} {\bf C37} (1988) 477;
A. Margaritis, G. Odor and A. Patkos, {\it ibid.} {\bf C39} (1988) 109;
M. Alford, {\it Phys. Lett.} {\bf B336} (1994) 237;
R. D. Ball, P. E. Haagensen, J. I. Latorre and
E. Moreno, {\it ibid.} {\bf B347} (1995) 80;
R. D. Ball and R. S. Thorne, {\it Ann. Phys.} {\bf 236} (1994) 117.
\medskip
\item{\paris} J. Polonyi, in the Proceedings of the Workshop on
``QCD Vacuum Structure and Its Applications'', edited by H.M. Fried and
B. M\"uller (World Scientific, Singapore, 1993), p. 3.
\medskip
\item{\fraser} C. M. Fraser, {\it Z. Phys.} {\bf C28} (1985) 101;
I. Aitchison and C. Fraser, {\it Phys. Rev.} {\bf D32} (1985) 2190.
\medskip
\item{\ilawrie} see for example, I. D. Lawrie, {\it J. Phys.}
{\bf A26} (1993) 6825; A. M. Nemirovsky and K. F. Freed, {\it ibid.}
{\bf A19} (1989) 591.
\medskip
\item{\morris} T. R. Morris, {\it Int. J. Mod. Phys.} {\bf A9}
(1994) 2411; {\it Phys.Lett.} {\bf B334}
(1994) 355 and {\bf B329} (1994) 241; {\it Nucl. Phys. } {\bf B458} (1996) 477.
\medskip
\item{\rivers} see, for example, R. J. Rivers, {\it Path Integral Methods 
in Quantum Field Theory}, (Cambridge, Cambridge, 1984); H. Kleinert, {\it
Gauge Fields in Condensed Matter}, vol.1, (World Scientific, 1989). 
\medskip
\item{\ginsparg} P. Ginsparg, {\it Nucl. Phys.} {\bf B170} (1980) 388;
T. Appelquist and J. Carazzone, {\it Phys. Rev.} {\bf D11} (1975) 2856.
\medskip
\item{\landsman} E. L. M. Koopman and N. P. Landsman,
{\it Phys. Lett.} {\bf B223} (1989) 421; N. P. Landsman, {\it Nucl. Phys.}
{\bf B322} (1989) 498; D. Gross, R. Pisarski and A. Yaffe, 
{\it Rev. Mod. Phys.} {\bf 53} (1981) 43.
\medskip
\item{\wetterich} see for example, N. Tetradis and C. Wetterich,
{\it Int. J. Mod. Phys.} {\bf A9} (1994) 4029, and {\it Nucl. Phys.}
{\bf B422} (1994) 541.  
\medskip
\item{\fiveloop} H. Kleinert , J. Neu , V. Schulte-Frohlinde, 
K.G. Chetyrkin, S.A. Larin, {\it Phys. Lett.} {\bf B272}, (1991) 39.
\medskip
\item{\dattanasio} M. D'Attanasio and M. Pietroni, 
{\it Nucl. Phys. } {\bf B472} (1996) 711.
\medskip
\item{\largen} D. O'Connor, C. R. Stephens and A. J. Bray, cond-mat/9601146;
M. Reuter, N. Tetradis and C. Wetterich,
{\it Nucl. Phys.} {\bf B401} (1993) 567.
\medskip
\item{\suranyi} J. A. Hertz, {\it Phys. Rev.} {\bf B14} (1976) 1165;
H. Hamidian, G. Semenoff, P. Suranyi, and L.C.R. Wijewardhana,
{\it Phys. Rev. Lett} {\bf 74} (1995) 4976.
\medskip
\item{\comellas} J. Comellas, Y. Kubyshin and E. Moreno, 
hep-th/9512086 and hep-th/9601112.
\medskip
\item{\gauge} S.-B. Liao, {\it Phys. Rev.} {\bf D53} (1996) 2020.
\medskip
\item{\wkrg} see, for example, M. Bonini, M. D'Attanasio and G. Marchesini,
{\it Nucl. Phys.} {\bf B437} (1995) 163 and {\bf B409} (1993) 441;
M. Reuter and C. Wetterich, 
{\it ibid.} {\bf B417} (1994) 181.
\medskip

%
%
%
%
%
%
%

\vfill
\break

\centerline{\bf FIGURE CAPTIONS}
\bigskip

\itemitem{Figure 1.}
Temperature dependence of the wavefunction renormalization constant
${\cal Z}_{\beta,k=0}(\Phi)$. Inset shows the  $\Phi$ dependence of 
${\cal Z}_{\beta,k=0}(\Phi)$ at $T=0$.
\medskip
\itemitem{Figure 2.}
Blocked potential $U_{\beta,k}(\Phi)$ near $T_c$. The transition is second
order. 
\medskip
\itemitem{Figure 3.}
Schematic diagram of the two methods used for measuring the critical exponents.
The left block summarizes method (1) and the right method (2).
\medskip
\itemitem{Figure 4.}
Effective critical exponent $\gamma_{\rm eff}$
as function of (a) ${\rm log}_{10}(T - T_c(k))$  and (b) 
${\rm log}_{10}((T - T_c(k))/k^{1/\nu_{\rm eff}})$, 
where the three-dimensional
value $\nu_3^{-1}=\nu^{-1}_{\rm eff}(0)=1.49 \pm 0.01$ 
is obtained via measuring the shift $|T_c-T_c(k)|$ due to the 
finiteness of $k$.
The different symbols correspond to different values of 
$k$.
\medskip
\itemitem{Figure 5.}
Dependence of $\gamma_{\rm eff}$ on 
$\bar\beta_c$ obtained via polynomial expansion of 
$U_{\beta,k}(\Phi)$ up to
$O(\Phi^8)$. The values of $\gamma_3$ and 
$\gamma_4$ calculated from  
integrating the RG improved versions of (3.1) and (3.2) are also included.
\medskip
\itemitem{Figure 6.}
Dependence of the minimum , $\hat\Phi_{\beta,k}$, on the
IR cutoff $k$ at the critical temperature
$T=T_c(k=0)$, showing that $\hat\Phi_{\beta,k}$ only
vanishes when $k = 0$.
\medskip
\itemitem{Figure 7.}
Temperature dependence of the thermal mass parameter
$\mu^2_{\beta}$, coupling constant 
$\lambda_{\beta}$, and 
the minimum $\hat\Phi_{\beta,k}$ near 
$T=T_c$. The $T$
dependence of $\lambda_{\beta}$ in the symmetric phase is also 
illustrated.
\medskip
%
\itemitem{Figure 8.}
$U^{(4)}_{\beta,k}$ as a function of $\Phi$ close to $T_c$ 
($T - T_c(k)\sim 10^{-5}$) calculated with the nontruncated RG equations.

\bigskip
\centerline{\bf TABLES}
\bigskip

\itemitem{Table 1.}
Critical exponents as function of the level of truncation along with the best 
calculations to date and experimental values.  NT indicates results obtained
from the nontruncated RG equations \rgft\ and \rgz.
\medskip
\itemitem{Table 2.}
$k$ dependence of vertex functions at $T_c$ in the absence of  
wavefunction renormalization.
\end